\documentclass[aps, pra, reprint, superscriptaddress,longbibliography]{revtex4-1}
\usepackage[utf8]{inputenc}
\usepackage{amsmath}
\usepackage{amssymb}
\usepackage{hyperref}
\usepackage{graphicx}
\usepackage{epstopdf}
\usepackage{dcolumn}
\usepackage{bm}
\usepackage{bbm}
\usepackage{xspace}
\usepackage{verbatim}
\usepackage{color}

\newcommand{\red}[1]{\textcolor{red}{#1}}

\DeclareMathOperator{\Tr}{Tr}

\hypersetup{colorlinks=true,linkcolor=blue,citecolor=blue, filecolor=blue,urlcolor=blue,breaklinks=true}

\begin{document}
\title{Probing of nonlinear hybrid optomechanical systems via partial accessibility}

\author{V. Montenegro}
\email{vmontenegro@uestc.edu.cn}
\affiliation{Institute of Fundamental and Frontier Sciences, University of Electronic Science and Technology of China, Chengdu 610051, China}

\author{M. G. Genoni}
\email{marco.genoni@fisica.unimi.it}
\affiliation{Quantum Technology Lab $\&$ Applied Quantum Mechanics Group, Dipartimento di Fisica {\em Aldo Pontremoli}, Universit\`a degli Studi di Milano, I-20133 Milano, Italia}

\author{A. Bayat}
\email{abolfazl.bayat@uestc.edu.cn}
\affiliation{Institute of Fundamental and Frontier Sciences, University of Electronic Science and Technology of China, Chengdu 610051, China}

\author{M. G. A. Paris}
\email{matteo.paris@fisica.unimi.it}
\affiliation{Quantum Technology Lab $\&$ Applied Quantum Mechanics Group, Dipartimento di Fisica {\em Aldo Pontremoli}, Universit\`a degli Studi di Milano, I-20133 Milano, Italia}

\date{\today}
\begin{abstract}
Hybrid optomechanical systems are emerging as a fruitful architecture for quantum technologies. Hence, determining the relevant atom-light and light-mechanics couplings is an essential task in such systems. The fingerprint of these couplings is left in the global state of the system during non-equilibrium dynamics. However, in practice, performing measurements on the entire system is not feasible, and thus, one has to rely on partial access to one of the subsystems, namely the atom, the light, or the mechanics. Here, we perform a comprehensive analysis to determine the optimal subsystem for probing the couplings. We find that if the light-mechanics coupling is known or irrelevant, depending on the range of the qubit-light coupling, then the optimal subsystem can be either light or the qubit. In other scenarios, e.g., simultaneous estimation of the couplings, the light is usually the optimal subsystem. This can be explained as light is the mediator between the other two subsystems. Finally, we show that the widely used homodyne detection can extract a fair fraction of the information about the couplings from the light degrees of freedom.
\end{abstract}

\maketitle

\section{Introduction}\label{sec:introduction}
The prime field of cavity quantum electrodynamics (QED)~\cite{Haroche:993568, Walther_2006, Mabuchi-science}, as a rich blend of atomic physics and quantum optics, has led to countless striking applications, including the one-atom maser~\cite{PhysRevLett.54.551}, one-atom laser~\cite{PhysRevLett.73.3375}, quantum gates~\cite{Reiserer2014, Hacker2016},  atom-cavity microscopy~\cite{Hood-ACM, Pinkse2000}, and novel quantum information and computation schemes~\cite{blais2004cavity, 10.2307/3559242}. Undoubtedly, one of the most fundamental model to investigate the coherent interplay between atom-field interactions is the Jaynes-Cummings model (JCM)~\cite{Greentree_2013}. As initially formulated, the JCM is composed of a two-level atom interacting with a quasi-resonant quantized cavity mode, being first employed to unravel the classical aspects of spontaneous emission~\cite{Shore93topicalreview}. To date, numerous extensions of the original model have been put forward, for instance, in the presence of multiple atoms~\cite{PhysRev.170.379, PhysRev.188.692} and arrays of coupled cavities~\cite{PhysRevA.76.031805, Greentree2006, Hartmann2006}.

While JCM involves interaction between an atom and a bosonic field, the field of cavity quantum optomechanics~\cite{RevModPhys.86.1391, 2015SCPMA..58.5648X, Yu-long} opens a new horizon by considering the interaction between two bosonic modes, namely a quantized electromagnetic field and a mesoscopic mechanical resonator. The canonical formulation of the optomechanical model considers a nonlinear coupling between the photon number and the position of the mechanical resonator~\cite{PhysRevA.51.2537}, making the physics of such systems inherently distinct from the linear (in the boson operator) nature of JCM. Similarly as for the JCM, the optomechanical systems have been also fully solved analytically for both time-dependent~\cite{Qvarfort_2020} and time-independent~\cite{PhysRevA.56.4175, PhysRevA.55.3042} Hamiltonians. Extensive experimental efforts and theoretical proposals have been devoted for the cooling of a mechanical object towards the ground state or even non-classical states~\cite{OConnell2010, Chan2011, Doherty2012PNAS,PhysRevX.8.041034, Liu_2013, PhysRevA.83.063835, Kleckner2006, Vanner2013, PhysRevLett.97.243905, PhysRevLett.92.075507, PhysRevLett.80.688, PhysRevB.69.125339, Jaehne_2008, PhysRevLett.117.077203, PhysRevA.98.053837,Kronwald2013,Wollman2015,Wilson2015,Pirkkalainen2015,Genoni2015,Genoni15, Rossi2018,Rossi19,Magrini2021,Tebbenjohanns2021}, which have led to a range of applications, such as quantum state transfer~\cite{PhysRevA.93.062339},   entanglement distillation~\cite{PhysRevA.100.042310}, quantum state engineering~\cite{PhysRevA.99.043836}, and quantum metrology purposes~\cite{Hosseini2014, PhysRevLett.97.133601, Chaste2012, Krause2012, Qvarfort2018, PhysRevResearch.3.013159, PhysRevResearch.2.043338}, to name a few.
\begin{figure}
\centering \includegraphics[width=\linewidth]{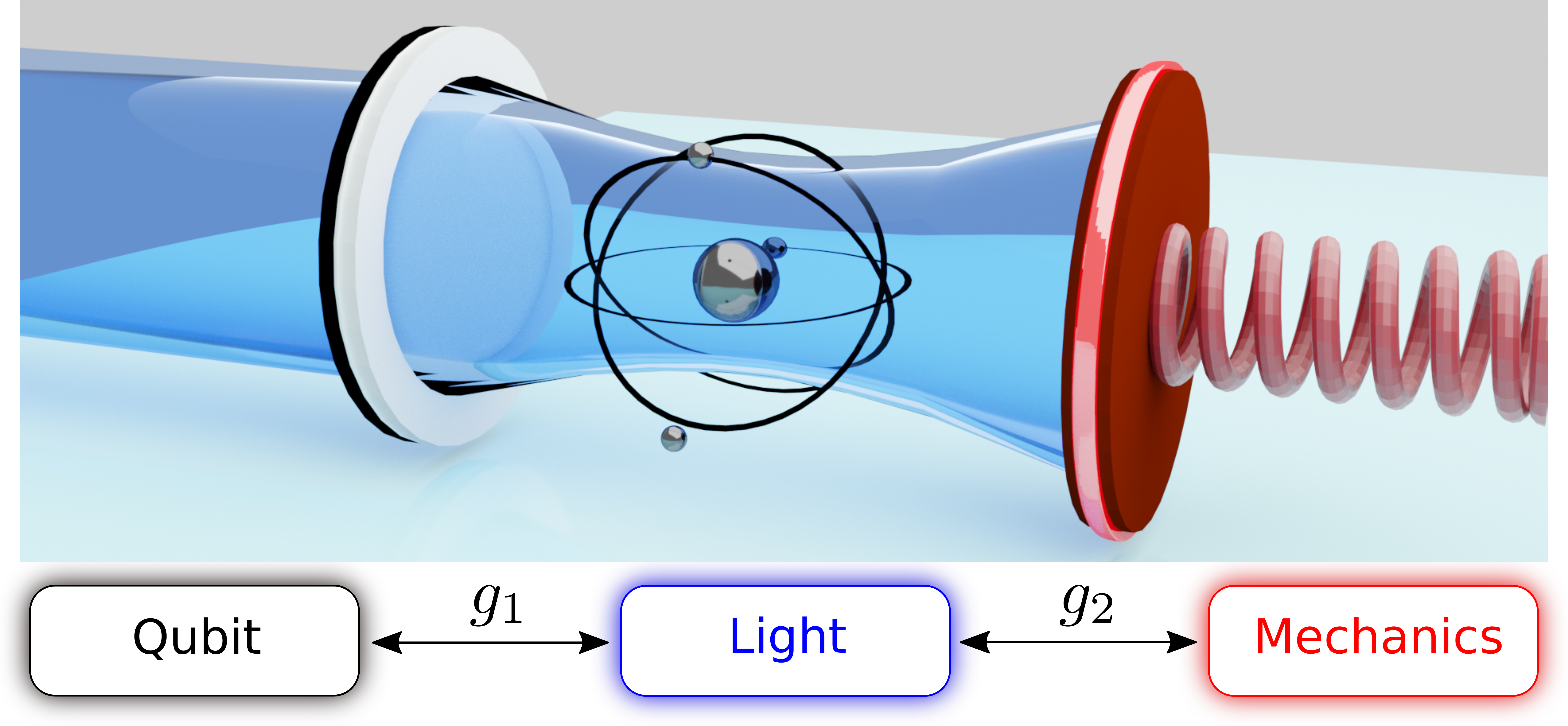}
\caption{Schematic of a hybrid optomechanical tripartite system. Two unknown parameters $g_1$ and $g_2$ are to be estimated in a cavity-mediated system with partial accessibility.}
\label{fig:sketch}
\end{figure}

Recently, cavity QED has been merged with optomechanics (see Fig.~\ref{fig:sketch}), enabling the emergence of vibrant novel hybrid systems~\cite{RogersLoGulloDeChiaraPalmaPaternostro+2014, Kurizki3866, Wallquist_2009, PhysRevLett.112.013601, PhysRevA.95.023832, PhysRevA.77.050307}. While the constituent elements of cavity QED, i.e., the atoms and the light field, operates near resonance, the building blocks of optomechanics, i.e., the light and the mechanical fields, are highly off-resonant. Therefore, the hybridization of quantum systems provide a rich playground for non-equilibrium dynamics involving distinct natural frequencies~\cite{e23080966, RogersLoGulloDeChiaraPalmaPaternostro+2014}. These versatile systems provide a fruitful architecture for various tasks for quantum technologies, including mechanical ground state cooling~\cite{Carmele_2014, _ernot_k_2019, PhysRevA.80.061803, Zeng2017, Nie:15, PhysRevA.94.033809, PhysRevLett.112.013601}, transducers for long-distance quantum communication~\cite{PhysRevLett.105.220501, PhysRevA.84.042341, Stannigel_2012, Habraken_2012, 10.1093/nsr/nwv048, PhysRevA.94.033809}, mechanical non-classical state preparation~\cite{Bergholm_2019, Carmele_2014}, tunable photon blockade effects~\cite{PhysRevA.92.033806}, quantum entanglement~\cite{PhysRevA.77.050307}, and quantum nonlocality~\cite{Zhangnonlocality}. Thanks to recent advancements in quantum technologies, these structures have been proposed and realized in various physical platforms, resulting in a different range of couplings~\cite{xiang2013hybrid}. 

Consequently, the precise estimation of the couplings between the atom-light and the light-mechanics is a key step for harnessing the hybrid systems for practical applications. In order to approach this particular problem, we will then exploit the tools given by local quantum estimation theory (QET)~\cite{helstrom1976quantum,Holevo2011b,Paris2009,PerspectiveMultiPar,Liu_2019,Yuan2017} (for a global quantum sensing scheme see, e.g., Ref.~\cite{PhysRevLett.126.200501}), whose aim is indeed to provide the ultimate bounds on parameter estimation in quantum systems, and to assess the usefulness of practical measurement strategies. This kind of analysis has already been done for the estimation of Hamiltonian coupling constants, both for light-matter interactions \cite{MarcoCarmenJC, BernardJCEst,PhysRevA.99.032122} and for optomechanical systems \cite{LatmiralOptoEst,BernardOptoEst,SanavioOptoEst,SalaOptoEst}. In general, as the system evolves, the information of the relevant couplings is imprinted in the quantum state of the global system~\cite{PhysRevLett.115.110401, PhysRevLett.126.070503}. To extract such information and estimate the parameters of interest, one must perform an appropriate measurement on the system. In practice, however, the accessibility to the whole system for performing a global measurement is unlikely. Therefore, one has to resort to partial accessibility in which only one subsystem can be measured for inferring the information about the system. Since the dynamics make the subsystems of these hybrid structures highly entangled, the information contained in each of these subsystems is smaller than the global state. Indeed, quantum many-body sensors with partial accessibility show reduced sensitivity in spin chains, demanding complex driving necessary for restoring the precision~\cite{PhysRevLett.127.080504, mishra2021integrable}. Thus, several issues should be addressed for probing the coupling in hybrid optomechanical systems by partial accessibility. First, how the information of the couplings spreads between the subsystems and how much of it can be extracted with partial accessibility. Second, which subsystem has the maximum information content, and thus, is the best to be measured for estimating the couplings. Third, by only considering the practically available measurements, what fraction of the information content can be experimentally extracted.  

This paper addresses the above issues in a hybrid optomechanical system composed of a two-level atom, a cavity field, and a mechanical oscillator. The goal is to determine the couplings between the atom-light and the light-mechanics over a wide range, considering only partial accessibility to one of the subsystems. We consider three different regimes: (i) estimating one coupling while the other is known (single parameter estimation); (ii) estimating one coupling while the other is unknown (parameter estimation with nuisance parameters); and (iii) estimating both of the couplings simultaneously (joint estimation). For estimating the atom-light coupling, we found that in the single- and nuisance multi-parameter estimation, depending on the range of couplings, either light or atom can be the optimal subsystem to be measured. In all other cases, the light is the dominant optimal subsystem to be measured. Surprisingly, measuring the mechanical degrees of freedom is hardly helpful for estimating the couplings. This can be understood as the light mediates the interaction between the other subsystems, thus carrying most of the information. Nonetheless, the optimal measurement basis on light degrees of freedom is very complex. Thus, we focus on the widely available homodyne detection for the estimation of the couplings. Our analysis shows this can indeed determine the couplings simultaneously with fair precision.

The rest of the paper is organized as follows: In Sec.~\ref{sec:parameter-estimation} we present preliminaries on quantum parameter estimation. In Sec.~\ref{sec:the-model} we introduce the hybrid optomechanical model and a brief analysis on the entanglement dynamics. In Sec.~\ref{sec:single-parameter estimation}, we address the single-parameter estimation scenario and the optimal subsystem which provides the more information content. We present the multi-parameter case in Sec.~\ref{sec:multi-parameter estimation}, including nuisance and joint estimation protocols. We also investigate how much information one can extract with available homodyne detection schemes. Finally, we conclude our work in Sec.~\ref{sec:conclusions}. Two appendices are included in Secs.~\ref{appendix-1} and ~\ref{appendix-2}, involving details about the hybrid optomechanical system and the inclusion of decoherence, respectively.

\section{Bits of quantum parameter estimation}\label{sec:parameter-estimation}
In this section we will provide the basic ingredients of local quantum estimation theory (we refer to \cite{helstrom1976quantum,Holevo2011b,Paris2009,PerspectiveMultiPar} for more details and explanations). We will start by considering the single parameter case and we will then extend the formalism to the multiparameter one. We thus consider a family of states $\varrho_\lambda$, where $\lambda$ is the parameter that one wants to estimate.  In a quantum mechanical setting, one performs a measurement described by a positive-operator valued measure (POVM) $\{\Pi_x\}$, such that the whole process is described by the conditional probability 
\begin{align}
p(x|{\lambda}) = \Tr[\rho_{{\lambda}} \Pi_x] \,.
\end{align}
After obtaining a statistical sample of $M$ outcomes $\mathcal{X}=\{x_1,\dots, x_M\}$, one can then define an estimator $\tilde{{\lambda}}(\mathcal{X})$  to infer the value of $\lambda$. The variance of any unbiased estimator, that is such that $\mathbbm{E}[\tilde{{\lambda}}(\mathcal{X})] = \lambda$, is proven to be bounded according to the Cram\'er-Rao bound
\begin{align}
{{\rm Var}}({\lambda}) \geq \frac{1}{M F} \,, \label{eq:singleCRB}
\end{align}
where we have introduced the (classical) Fisher information
\begin{align}
{F} = \int dx \frac{1}{p(x|\lambda)}\left( \frac{\partial p(x|\lambda)}{\partial \lambda} \right)^2 \,. \label{eq:classical_fisher}
\end{align}
This bound can be in principle saturated via optimal estimators, such as the maximum likelihood or the Bayesian estimator. In quantum mechanics is then possible to define a more general bound, that depends only on the quantum statistical model $\rho_{{\lambda}}$, and not on the particular measurement performed $\{\Pi_x\}$. In particular one proves the quantum Cramer-Rao bound
\begin{align}
{\rm Var}({\lambda}) \geq \frac{1}{M F}  \geq \frac{1}{M Q}  \,, \label{eq:singleQCRB}
\end{align}
where we can now define the quantum Fisher information (QFI) 
\begin{align}
{Q} = \Tr [ \rho_{{\lambda}} L_\lambda^2 ] \,
\end{align}
with the symmetric logarithmic derivative (SLD) operator $L_\lambda$ implicitly defined by the Lyapunov equation 
\begin{align}
\frac{\partial\rho_{{\lambda}}}{\partial{\lambda}} = \frac{L_\lambda \rho_{{\lambda}} + \rho_{{\lambda}} L_\lambda}{2} \,.
\label{eq:SLD}
\end{align}
Remarkably one can prove that the bound (\ref{eq:singleQCRB}) can be always saturated, that is one can always find an optimal POVM $\{\Pi_x\}$, such that the corresponding classical Fisher information $F$ is equal to the QFI $Q$. 
 
In the multiparameter scenario, the family of quantum states $\rho_{\boldsymbol{\lambda}}$ is defined in terms of a set of unknown $d$ parameters $\boldsymbol{\lambda} = \{\lambda_1,\dots, \lambda_d\}$. The bounds (\ref{eq:singleCRB}) and (\ref{eq:singleQCRB}) can be generalized as matrix inequalities for the covariance matrix ${\bf {\rm Cov}}(\boldsymbol{\lambda})$ of any unbiased estimatior as 
\begin{align}
{\bf {\rm Cov}}(\boldsymbol{\lambda}) \geq \frac{1}{M} \mathcal{F}^{-1}  \geq \frac{1}{M} \mathcal{Q}^{-1} \,, \label{eq:matrixQCRB}
\end{align}
where we have introduced the classical and quantum Fisher information matrices with elements
\begin{align}
\mathcal{F}_{ij} &= \int dx \frac{1}{p(x|\boldsymbol{\lambda})} \left( \frac{\partial p(x|\boldsymbol{\lambda})}{\partial \lambda_i }\right)\left( \frac{\partial p(x|\boldsymbol{\lambda})}{\partial \lambda_j }  \right) \,, \label{eq:classical_fisher_matrix} \\
\mathcal{Q}_{ij} &= \Tr\left[ \rho_{\boldsymbol{\lambda}} \frac{L_i L_j + L_j L_i}{2} \right]  \label{eq:quantum_fisher_matrix} \,,
\end{align}
and where one defines a different SLD operator $L_j$ for each parameter $\lambda_j$. The matrix bounds above can be translated into a family of scalar bounds, where in particular we will consider the one for the sum of the variances of each parameter
\begin{align}
\sum_j {\rm Var}(\lambda_j) \geq \frac{1}{M} \Tr[ \mathcal{F}^{-1}] \geq \frac{1}{M} \Tr[\mathcal{Q}^{-1}] \,. \label{eq:scalarQCRB}
\end{align}
One of the main differences between the single- and the multi-parameter scenario is that, while the classical Cram\'er-Rao bounds (both matrix and scalar) defined in terms of the classical FI matrix $\mathcal{F}$ can be in principle saturated, this is not in general the case for the quantum Cram\'er-Rao bounds dictated by the QFI matrix $\mathcal{Q}$. This fact can be understood by observing that in general optimal measurements for different parameters may correspond to non-commuting observables. This led to the formulation of several other bounds that may be more tight under certain conditions~\cite{PerspectiveMultiPar}. However in this work we will not focus on this aspect, and we will rather consider the scalar bound (\ref{eq:scalarQCRB}) as an ultimate benchmark able to give relevant  information on the multiparameter estimation properties of the quantum system under exam, and we will then focus on a particular feasible measurement strategy and to the corresponding (pontentially achievable) classical bound.

In the framework of multiparameter quantum estimation, falls also the case of {\em nuisance quantum estimation} \cite{SuzukiNuisance}: suppose we are interested only in a single parameter $\lambda_j$ from the set of $d$ unknown parameters $\boldsymbol{\lambda}$. In this case, the other $d-1$ parameters are typically called {\em nuisance parameters}, and the bound on the variance of any estimator of the parameter $\lambda_j$ reads
\begin{align}
{\rm Var}({\lambda}_j) \geq \frac{1}{M} (\mathcal{F}^{-1})_{jj}  \geq \frac{1}{M} (\mathcal{Q}^{-1})_{jj}  \,, \label{eq:nuisanceQCRB}
\end{align}
where the inverse of the diagonal elements of the classical and quantum Fisher information $F=\mathcal{F}_{jj}$ and $Q=\mathcal{Q}_{jj}$ in Eq. (\ref{eq:singleQCRB}) have been replaced by the diagonal elements of the corresponding inverse matrices. One has that in general $(\mathcal{F}^{-1})_{jj} \geq (\mathcal{F}_{jj})^{-1}$ and $(\mathcal{Q}^{-1})_{jj} \geq (\mathcal{Q}_{jj})^{-1}$, confirming the fact that having less information on the other parameters can only lead to a worse estimation of the parameter $\lambda_j$. We however remark that in this case, the ultimate bound (\ref{eq:nuisanceQCRB}) for a single parameter $\lambda_j$ can be in principle achieved, with the optimal measurement strategy that coincides with the one that is optimal in the {\em nuisance-free} scenario.

As a technical remark, we point out that in order to derive the bounds described in this section, it is necessary to know the derivative of the operator respect to the parameters $\boldsymbol{\lambda}$, for example in order to find the SLD operators as in Eq. (\ref{eq:SLD}). In our case, we will need to often resort to numerical procedures in order to evaluate this derivative. In particular we will compute the five-point stencil first derivative approximation with respect to $\lambda_i$ and increment $\Delta \lambda_i \ll 1$:
\begin{eqnarray}
\nonumber \frac{\partial f(\lambda_i)}{\partial\lambda_i} &\approx& [-f(\lambda_i+2\Delta \lambda_i)+8f(\lambda_i+\Delta \lambda_i)\\
&-&8f(\lambda_i-\Delta \lambda_i)+f(\lambda_i-2\Delta \lambda_i)]/(12\Delta \lambda_i)\,, \label{eq:derivative}
\end{eqnarray}
which has an error of order $(\Delta \lambda_i)^4$.

\section{The model}\label{sec:the-model}
We consider a hybrid system composed of a two-level atom (qubit), a single electromagnetic (cavity) mode, and a (mechanical) harmonic oscillator. The qubit interacts with the cavity mode via Jaynes-Cummings Hamiltonian, whereas the cavity field couples to the mechanical oscillator through nonlinear optomechanical interaction~\cite{PhysRevLett.112.013601}. Indeed, the qubit and the mechanical parties will interact with the cavity mode undergoing entirely different Hamiltonians. The total cavity-mediated tripartite Hamiltonian is ($\hbar = 1$)
\begin{eqnarray}
\nonumber H &=& \omega_c a^\dagger a + \omega_m b^\dagger b + \frac{\omega_q}{2} \sigma_z + g_1 (\sigma^+a + \sigma^-a^\dagger)\\
&-& g_2 a^\dagger a (b^\dagger + b),\label{eq:H}
\end{eqnarray}
where the cavity (mechanical) mode is described by the bosonic operators satisfying $[a, a^\dagger]=\mathbb{I}$ ($[b, b^\dagger]=\mathbb{I}$) with natural frequency $\omega_c$ ($\omega_m$). The qubit is described by Pauli matrices $\sigma_{x,y,z}, \sigma^+=|e\rangle \langle g|, \sigma^-=|g\rangle \langle e|$ with energy gap $\omega_q$ between the ground state $|g\rangle$ and the excited energy level $|e\rangle$. The Jaynes-Cummings interaction term, $\sigma^+a + \sigma^-a^\dagger$, accounts for the annihilation (creation) of a photonic excitation in the cavity by (de-)exciting the qubit ground state (excited state) with coupling strength $g_1$. The nonlinear optomechanical Hamiltonian, $-g_2 a^\dagger a (b^\dagger + b)$, couples the cavity number operator directly to the mechanical object's position $(\propto{(b^\dagger + b)})$ via radiation-pressure interaction with strength $g_2$. We would like to estimate the coupling parameters $g_1$ and $g_2$ through the dynamics of the system when the accessibility to the system is limited, such that only one part of the hybrid system can be measured.

The Jaynes-Cummings Hamiltonian assumes $|\omega_q - \omega_c|\ll \omega_q + \omega_c$ for its derivation, which in turn allows us to neglect the fast temporal oscillations while keeping the rotating terms $\sigma^+a$ and $\sigma^-a^\dagger$~\cite{nla.cat-vn674977}. As it is known, this rotating wave approximation holds valid when $g_1 \lesssim 0.1\omega_c$. Going beyond this regime, one necessarily needs to describe the dynamics with the isotropic quantum Rabi model as the counter-rotating terms $\sigma^+a^\dagger$ and $\sigma^-a$ give rise to experimentally measurable effects~\cite{Niemczyk2010}. The single-photon optomechanical coupling $g_2$ highly vary depending of the experimental setup considered~\cite{Qvarfort2018}. Nonetheless, in the nonlinear regime its value range typically from $g_2 \ll \omega_m$ to $g_2 \lesssim 0.2 \omega_m$ for certain novel architectures~\cite{Chan2011, Murch2008, PhysRevLett.109.223601, Kaviani:15, Vanner16182, PhysRevA.94.023841}. In general $\omega_m\ll \omega_c$ and in what follows, we assume $\omega_c=\omega_q$ and $\omega_c = 100\omega_m$. Based on these, we consider the coupling $g_1$ and $g_2$ varying within a range $g_1 \in (0,0.2\omega_m]$, and $g_2 \in (0,0.2\omega_m]$.

We commence by considering the closed system evolution (the open quantum case is studied in Appendix~\ref{appendix-2}) from the initial state:
\begin{equation}
|\psi(0)\rangle = |g\rangle\otimes|\alpha\rangle\otimes|\beta\rangle.\label{eq:initial-state-pure}
\end{equation}
Here, the qubit initializes in its ground state energy, whereas the cavity (mechanical) field evolves from a coherent state of amplitude $\alpha=2$ ($\beta=2$). The system evolves under the action of the Hamiltonian as $|\psi(t)\rangle=e^{-iHt}|\psi(0)\rangle$. The quantum state of each subsystem is described by a reduced density matrix through tracing out the other parties
\begin{equation}
\rho_{s}(t)=\mathrm{Tr}_{\hat{s}}[|\psi(t)\rangle\langle\psi(t)|],
\end{equation}
where the label $s$ accounts for the cavity, qubit, or mechanical subsystems and $\mathrm{Tr}_{\hat{s}}[\cdot]$ means tracing out the complementary parts of the subsystem $s$. As the system evolves, the information of $g_1$ and $g_2$ is imprinted in the wave function $|\psi(t)\rangle$, and thus, $\rho_s(t)$. Regarding the estimation of $g_1$ and $g_2$, one can raise two open questions: (i) which subsystem is more informative about the values of $g_1$ and $g_2$; and (ii) what percentage of the global information content, extracted from the global state $|\psi(t)\rangle$, can be accessed through each subsystems. In the following sections, we address these issues.

\subsection{Entanglement dynamics}
The evolution governed by the Hamiltonian in Eq.~\eqref{eq:H} is complex and can be only solved analytically in the limit of vanishingly small $g_2$, namely $g_2\ll\omega_m$. Therefore, for a general case we have to compute the evolution numerically. To understand the dynamics of the system, we investigate how each subsystem entangles in time, outlining relevant remarks on the tripartite correlation dynamics. To do so, we compute the von Neumann entropy
\begin{equation}
S(t)=-\mathrm{Tr}\left[\rho_s(t) \mathrm{log}_2 \rho_s(t)\right],
\end{equation}
which quantifies the degree of entanglement between the subsystem $s$ and the rest of the system. Here, the logarithm in base 2 sets the upper limit for a maximally entangled qubit subsystem as 1. Furthermore, we numerically truncate the cavity and the mechanical parties up to $n=25$ of bosons, and hence one can fairly compare the entanglement between such subsystems.
\begin{figure}[t]
\includegraphics[width=\linewidth]{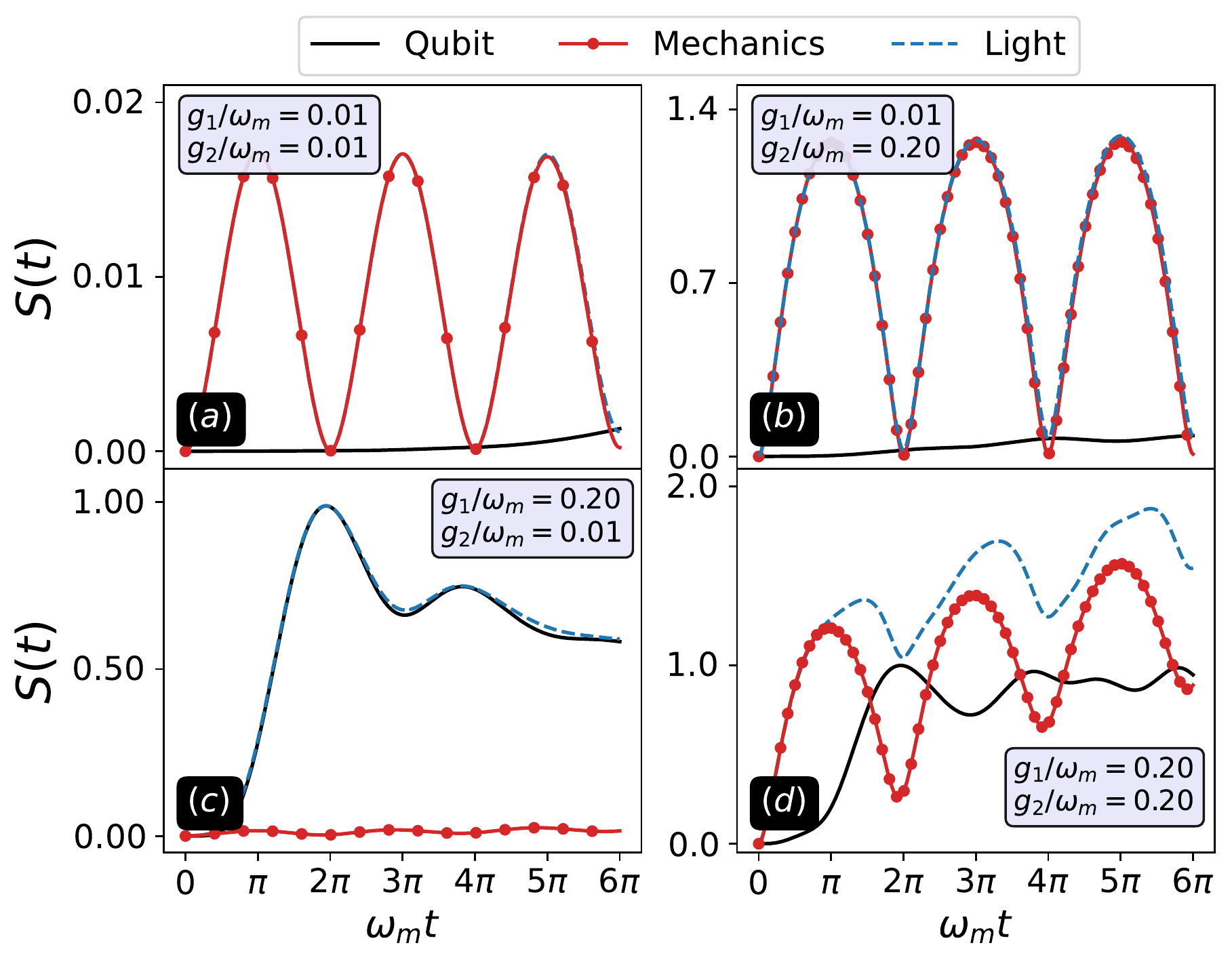}
\caption{Von Neumann entropy $S(t)$ for each subsystem as a function of the scaled time $\omega_m t$ for different $g_1$ and $g_2$ coupling parameters. The dynamics evolves from an initial state as in Eq.~\eqref{eq:initial-state-pure}.}
\label{fig:vn-entropy}
\end{figure}
In Fig.~\ref{fig:vn-entropy} we compute the von Neumann entropy $S(t)$ for each subsystem as a function of the scaled time $\omega_m t$ for four representative $g_1$ and $g_2$ coupling parameters. In Fig.~\ref{fig:vn-entropy}(a) we consider a regime where both $g_1$ and $g_2$ are small, namely $g_1=g_2=0.01\omega_m$. As the figure shows, each party weakly entangles with the rest of the system. However, a noticeable oscillatory entanglement of both the cavity and the mechanical oscillator within each mechanical cycle takes place. This is because the optomechanical interaction occurs in time scales of the order of $1/\omega_m$, while the qubit-light interaction, here scaled by $\omega_m$, evolves in slower times proportional to $1/\omega_c$. In Fig.~\ref{fig:vn-entropy}(b), we consider the situation where $g_2\gg g_1$, namely $g_2=0.2\omega_m$ and $g_1=0.01\omega_m$. As evident from the figure, the nonlinear optomechanical evolution dominates over the almost negligible qubit entanglement, showing the coherent light-matter dynamics due to the well-known nonlinear Kerr-like coherent phase~\cite{PhysRevA.56.4175, RevModPhys.86.1391}. The entanglement of the cavity and the mechanical  oscillator takes its maximum at half of the mechanical oscillator's cycle~\cite{PhysRevA.56.4175}. Furthermore, one finds the expected mechanical disentanglement at multiples of $2\pi\omega_m t$, which due to the presence of the qubit, the disentanglement is only approximated. Note that in Figs.~\ref{fig:vn-entropy}(a) and (b), the entanglement of the cavity and the mechanical oscillator follow an almost identical curve. This is because in these two regimes $g_1$ is very small, and thus, the qubit is almost disentangled from the rest. Therefore, most of the entanglement is coming from the bipartite entanglement between the cavity and the mechanical oscillator. In Fig.~\ref{fig:vn-entropy}(c), we consider the opposite regime where $g_1\gg g_2$, namely $g_1=0.2\omega_m$ and $g_2=0.01\omega_m$. Interestingly, the qubit (almost overlapped with the light subsystem) entangles maximally with the rest of the system at the first mechanical oscillator, remaining considerably high within the time interval $\omega_m t\in (0, 6\pi]$. This is because in this regime $g_2$ is very small, and thus, the mechanical oscillator remains almost disentangled from the others. In Fig.~\ref{fig:vn-entropy}(d), we consider a regime where both $g_1$ and $g_2$ are not negligable, namely $g_1 = g_2 = 0.2\omega_m$. As opposed to the above cases, there is an evident interplay between parties, and thus, one cannot only approximate its entanglement dynamics via Jaynes-Cummings or optomechanical Hamiltonians alone. Unlike Fig.~\ref{fig:vn-entropy}(b), the stronger presence of the qubit makes the mechanical party remain entangled after one period. Additionally, it is observed that the qubit keeps highly entangled within the time window, whereas the mechanical and the light parties now entangle differently, with the cavity field reaching higher values of entanglement with the rest of the system. This analysis shows that, with partial accessibility, one has a rich playground for sensing $g_1$ and $g_2$ and depending on their values the most relevant subsystem may be different.

\section{single-parameter estimation}\label{sec:single-parameter estimation}
We here start to derive the different bounds on the estimation precision for the two coupling parameters. We will thus consider the evolved quantum state as our quantum statistical model $\rho_{\boldsymbol{g}}$, with the vector of parameters $\boldsymbol{g}=\{g_1, g_2\}$. \\
We first focus on setting the precision limits for estimating only one coupling parameter assuming the other one is known. As presented in Sec.~\ref{sec:parameter-estimation}, in this case the ultimate bound is given by Eq. (\ref{eq:singleQCRB}), and thus the expression that quantifies the above is $(\mathcal{Q}_{ii})^{-1}$, for $i = 1,2$, and where $\mathcal{Q}$ is the QFI matrix corresponding to $\rho_{\boldsymbol{\lambda}}$.

\begin{figure*}[t]
\includegraphics[scale=0.48]{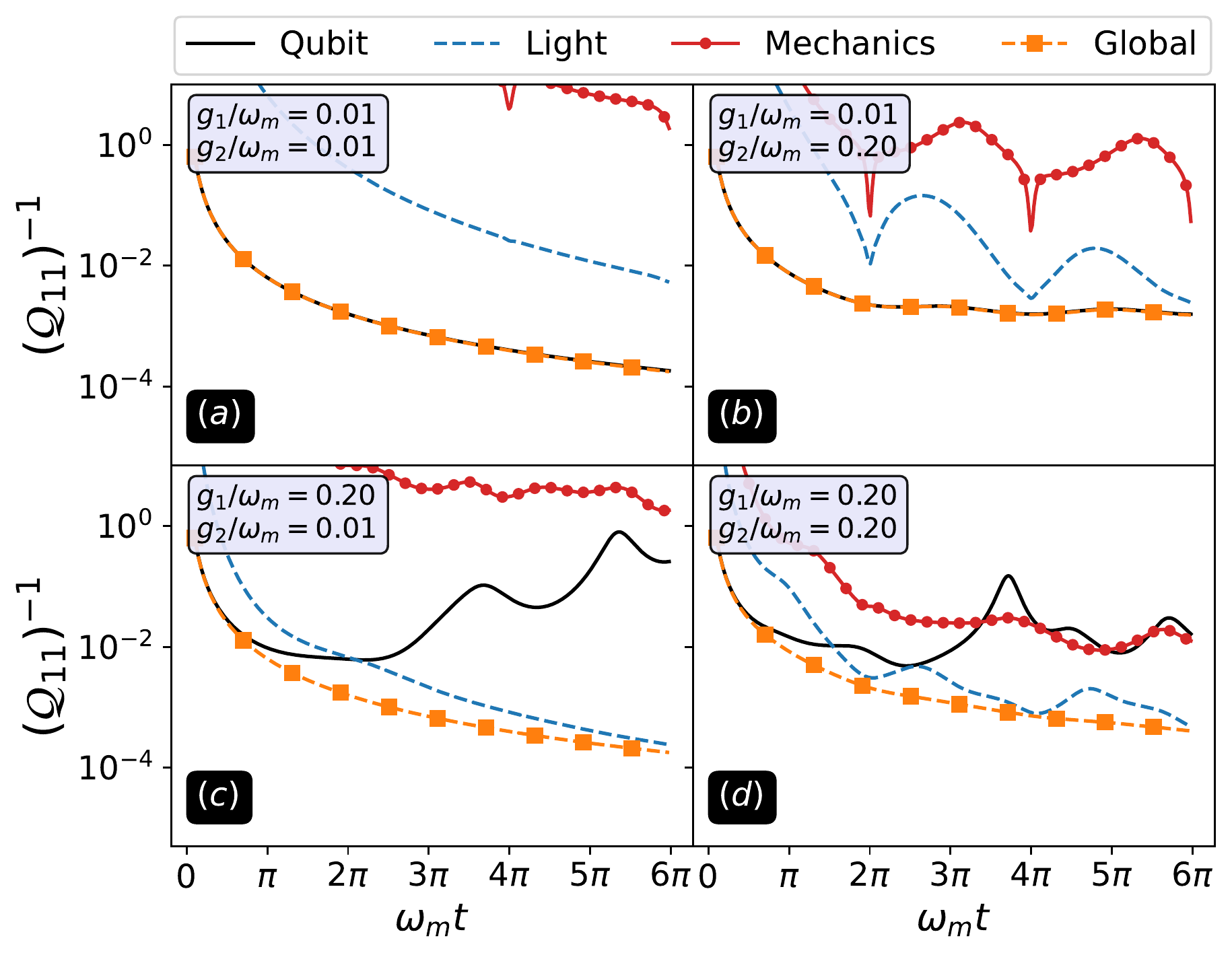}\includegraphics[scale=0.48]{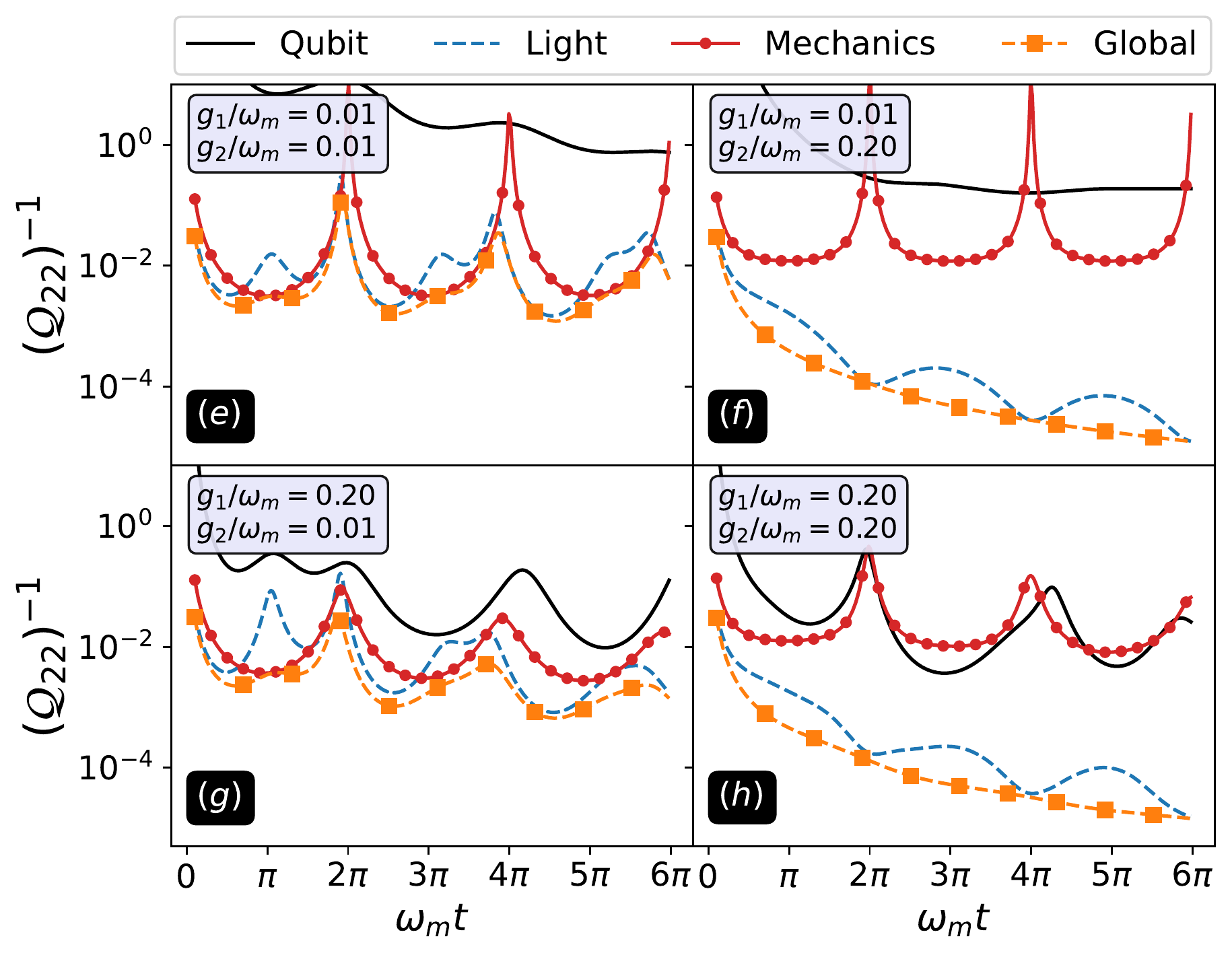}
\caption{Inverse of the quantum Fisher information $(\mathcal{Q}_{ii})^{-1}$ for partial access and for the global state for four coupling parameters. Panels (a) to (d) show the estimation of $g_1$ by knowing $g_2$; (e) to (h) show the estimation of $g_2$ by knowing $g_1$.}
\label{fig:single-parameter-estimation}
\end{figure*}

In Fig.~\ref{fig:single-parameter-estimation} we numerically evaluate the inverse of the quantum Fisher information $(\mathcal{Q}_{ii})^{-1}$ for the partial and global states for four relevant coupling parameters. 

Let us first focus on panels Figs.~\ref{fig:single-parameter-estimation}(a) to (d), where we show the precision limits in estimating $g_1$ by knowing $g_2$. In Figs.~\ref{fig:single-parameter-estimation}(a) and ~\ref{fig:single-parameter-estimation}(c), we consider weak optomechanical coupling $g_2=0.01\omega_m$ while the qubit-light coupling takes the values $g_1=0.01\omega_m$ and $g_1=0.2\omega_m$, respectively. Notice that, while the quantum state can be derived as in Eq.~\eqref{eq:so-equation}, the quantum Fisher information expression remains intractable. Hence, we rely on numerical simulations with the five-point method derivative as in Eq.~\eqref{eq:derivative}. As expected, the mechanical oscillator which can encode $g_1$ only through the cavity field remains highly disentangled from the rest of the system, due to small $g_2$ and thus plays an irrelevant role in estimating $g_1$. Nonetheless, as seen from Fig.~\ref{fig:single-parameter-estimation}(a), having partial access to the qubit subsystem nearly saturates the ultimate global bound. As $g_1$ increases, as shown in Fig.~\ref{fig:single-parameter-estimation}(c), a transition between the qubit and the light parties occur at $\sim \omega_mt=2\pi$ for delivering the best partial estimation performance. Thanks to the regime of $g_2 \ll \omega_m$, the mechanical system can be neglected and the dynamics can be approximated by Jaynes-Cummings evolution. Indeed, an initial state $|\psi_\mathrm{JC}(0)\rangle = |g,\alpha\rangle$ evolves as
\begin{equation}
|\psi_\mathrm{JC}(t)\rangle = c_0|g,0\rangle + \sum_{n=1}^\infty c_n(g_1)|g,n\rangle + d_n(g_1)|e,n-1\rangle,\label{eq:jc-global-pure}
\end{equation}
where
\begin{eqnarray}
\nonumber c_0 &=& e^{-|\alpha|^2/2},\\
\nonumber c_n(g_1) &=& c_0\frac{\alpha^n}{\sqrt{n!}}e^{-in\omega_ct}\cos(\sqrt{n}g_1t),\\
d_n(g_1) &=& -ic_0\frac{\alpha^n}{\sqrt{n!}}e^{-in\omega_ct}\sin(\sqrt{n}g_1t).\label{eq:jc-coefficients}
\end{eqnarray}
The QFI for the Jaynes-Cummings global pure state in Eq.~\eqref{eq:jc-global-pure} can be derived analytically, yielding
\begin{equation}
(\mathcal{Q}_{11})^{-1}_\mathrm{global, JC} = 4|\alpha|^2t^2,
\end{equation}
which is almost equal to the precision achievable from the global state Fisher information, orange-squared line in Fig.~\ref{fig:single-parameter-estimation}. To observe that the global bound is nearly saturated by the qubit subsystem, one realizes that for weak qubit-light coupling $g_1=0.01\omega_m$ the system evolves approximately as
\begin{equation}
|\psi_\mathrm{JC}(t)\rangle\approx(|g\rangle + \theta(g_1)|e\rangle)\otimes|\alpha\rangle,\label{eq:approx-jc}
\end{equation}
with a $g_1$-dependent coefficient $\theta(g_1)$. Obviously, all the information of $g_1$ is encoded in the state of the qubit, makes it the most relevant subsystem to be used for sensing $g_1$. In short, the vanishing value of $g_2$ makes the mechanical part irrelevant and the small value of $g_1$ implies that the information is almost fully encoded in the state of the qubit. Note that Eq.~\eqref{eq:approx-jc} is only an approximation to have a qualitative understanding of the dynamics. For instance, although the cavity state in Eq.~\eqref{eq:approx-jc} looks independent of $g_1$, the Fig.~\ref{fig:single-parameter-estimation}(a) clearly shows that cavity subsystem carries information about $g_1$. On the other hand, increasing $g_1$, see Fig.~\ref{fig:single-parameter-estimation}(c), makes the Rabi coefficients $c_n(g_1)$ and $d_n(g_1)$ to dynamically encode the parameter $g_1$ into the cavity subsystem undergoing a more complex dynamics as shown in Eq.~\eqref{eq:jc-coefficients}. In Fig~\ref{fig:single-parameter-estimation}(b), we consider $g_1{=}0.01\omega_m$ and $g_2{=}0.2\omega_m$. Again, similar to Fig.~\ref{fig:single-parameter-estimation}(a), because $g_1$ is small the information content of the qubit almost matches with the global state. However, since $g_2$ is large the mechanical state cannot be ignored anymore and one can write
\begin{equation}
|\psi(t)\rangle\approx(|g\rangle + \theta(g_1)|e\rangle)\otimes|\psi_\mathrm{OM}(t)\rangle,
\end{equation}
where $|\psi_\mathrm{OM}(t)\rangle$ is the quantum state of the cavity and the mechanical oscillator evolved from $|\psi_\mathrm{OM}(0)\rangle{=}|\alpha,\beta\rangle$. By ignoring the Jaynes-Cummings Hamiltonian, one can show that
\begin{equation}
|\psi_\mathrm{OM}(t)\rangle = c_0\sum_{n=0}^\infty\frac{\alpha^n}{\sqrt{n!}}e^{ig_2^2/\omega_m^2n^2(\omega_mt-\sin(\omega_mt))}|n,\phi_n(t)\rangle,\label{eq:state-optomechanics}
\end{equation}
where the coherent mechanical state evolves as
\begin{equation}
|\phi_n(t)\rangle =|\beta e^{-i\omega_mt}+g_2/\omega_mn(1-e^{-it\omega_m})\rangle.\label{eq:mechanics-coherent}
\end{equation} 
Obviously, the above is only an approximation as Fig.~\ref{fig:single-parameter-estimation}(b) clearly shows that the cavity and the mechanical states carry some information about $g_1$, although this is not evident in Eq.~\eqref{eq:state-optomechanics}. In Fig.~\ref{fig:single-parameter-estimation}(d), we consider $g_1=g_2=0.2\omega_m$. No approximation can be cast for this scenario. Interestingly, the cavity field is the subsystem which contains most of the information of $g_1$ and its achievable precision nearly  reaches the ultimate global bound. Notice that, for this regime, the mechanical oscillator also carries significant information about the coupling $g_1$, which is mediated through the cavity subsystem.

We now focus in panels Figs.~\ref{fig:single-parameter-estimation}(e)\red{-}(h), where we examine the precision bounds in estimating $g_2$ by assuming we know $g_1$. In Figs.~\ref{fig:single-parameter-estimation}(e) and ~\ref{fig:single-parameter-estimation}(f), we consider $g_1\ll\omega_m$ with optomechanical values $g_2=0.01\omega_m$ and $g_2=0.2\omega_m$, respectively. In this case, the qubit system (which can encode $g_2$ only through the cavity field) plays an irrelevant role in the estimation of $g_2$ due to the very weak coupling $g_1$. In Figs.~\ref{fig:single-parameter-estimation}(e) and (f), the mechanics shows high peaks at $\omega_mt = 2\pi k$, for some integer $k$. This poor performance in estimating $g_2$ is because such system almost returns to its initial state [see Eq.~\eqref{eq:mechanics-coherent}] at those times,
\begin{equation}
|\phi_n(\omega_mt=2\pi k)\rangle \approx |\beta\rangle
\end{equation}
and therefore, it becomes almost completely independent of $g_2$. However, the mechanical system maximally entangles with the light field at $\omega_mt = k\pi$, which explains the constant lower bounds reached by the mechanics shown in Figs.~\ref{fig:single-parameter-estimation}(e) and (f). Notably, as $g_2$ increases the information content of the cavity field about $g_2$ reaches close to the global bound as evidenced in Fig.~\ref{fig:single-parameter-estimation}(f). Under the coarse optomechanical approximation, i.e., $g_1=0$, one can easily prove that accessing the light field coincides with the ultimate global bound at multiples of $\omega_mt = 2k\pi$, $k$ being an integer. This is because at those times, the mechanical oscillator decouples from the light field and thus all the information of $g_2$ is transferred to the phase of the pure, decoupled cavity state. Indeed, from Eq.~\eqref{eq:state-optomechanics}, one finds
\begin{equation}
(\mathcal{Q}_{22})^{-1}_\mathrm{global, opto} = (\mathcal{Q}_{22})^{-1}_\mathrm{light} = \frac{(|\alpha|g_2k\pi)^{-2}}{64(1 + 6|\alpha|^2 + 4|\alpha|^4)}.
\end{equation}
In Fig.~\ref{fig:single-parameter-estimation}(g), we consider $g_1=0.2\omega_m$ and $g_2\ll \omega_m$. Here, the strong presence of the qubit coupled to the cavity field prevents the mechanical oscillator to return to its original state, as evidenced in the attenuated peaks at $\omega_mt=2\pi k$. In the regime where both $g_1$ and $g_2$ are strong, exemplified in Fig.~\ref{fig:single-parameter-estimation}(h) with $g_1=g_2=0.2\omega_m$, accessing the cavity field delivers excellent performance in estimating $g_2$, almost saturating the global bound, in the presence of strong qubit-light interaction $g_1$.

\subsection{Optimal subsystem}
As discussed in the previous section, accessing different subsystems give different estimation performances for a time interval and a given set of parameters $g_1$ and $g_2$. One can quantified the performance of each subsystem by comparing the $(\mathcal{Q}_{ii})^{-1}_\mathrm{sub}$ with the precision obtainable from the global state, namely $(\mathcal{Q}_{ii})^{-1}_\mathrm{global}$. Therefore, we define the single-parameter efficiency ratio
\begin{equation}
\eta_i^{(\mathrm{single})} = \frac{(\mathcal{Q}_{ii})^{-1}_\mathrm{global}}{(\mathcal{Q}_{ii})^{-1}_\mathrm{sub}}\Bigg\rvert_{t=t^*},\label{eq:efficiency-ratio}
\end{equation}
where $t^*$ is the time where the optimal subsystem reaches its minimum within a given time interval. In general, $0 \leq \eta_i^{(\mathrm{single})} \leq 1$ and each subsystem which contains more information about $g_i$ results in higher values of $\eta_i^{(\mathrm{single})}$.

\begin{figure}
\includegraphics[width=0.5 \textwidth]{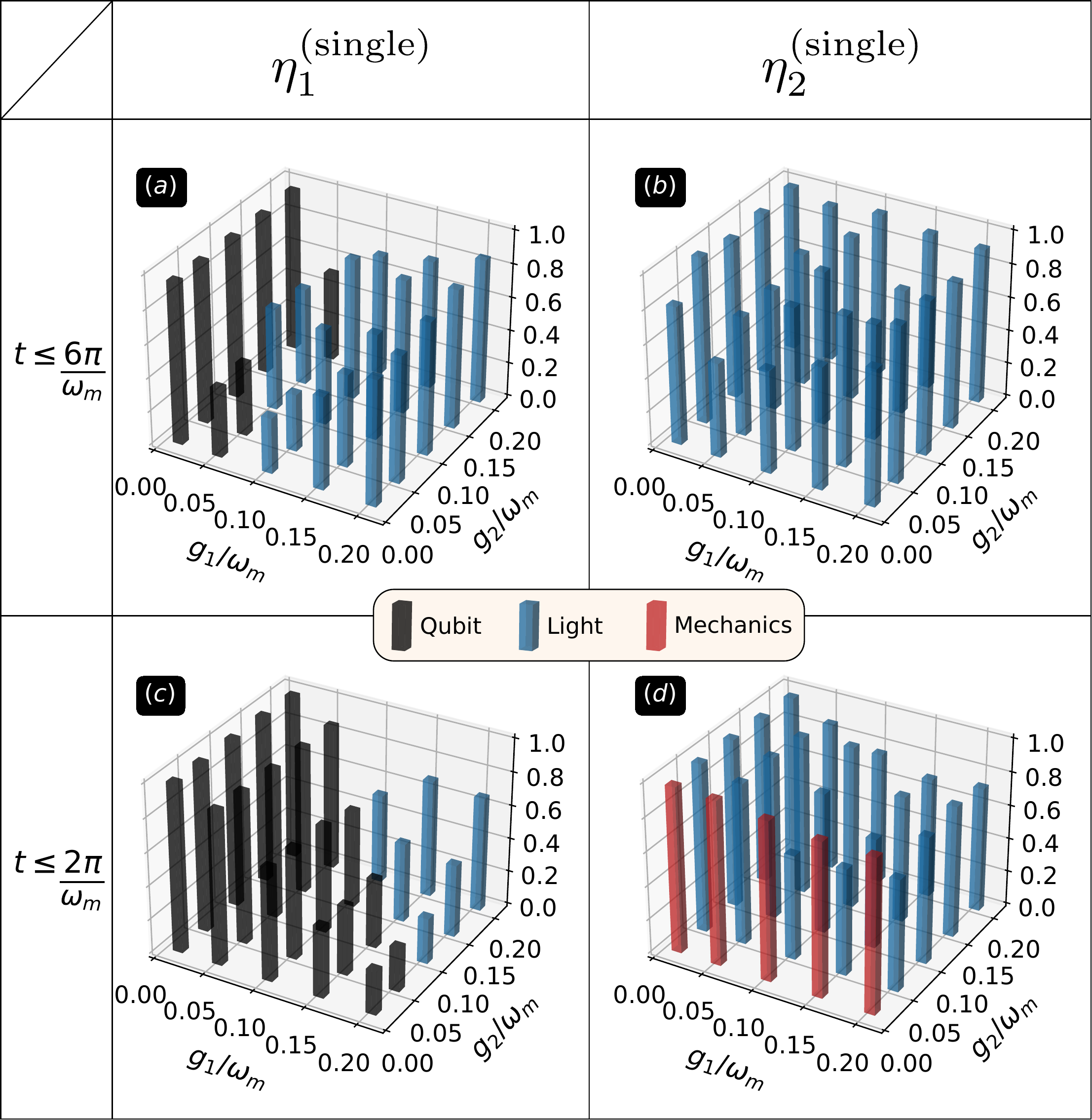}
\caption{Efficiency as defined as in Eq.~\eqref{eq:efficiency-ratio} for the estimation of $g_1$ [panels (a) and (c)] and $g_2$ [panels (b) and (d)] for two time intervals. The height of each bar quantifies the performance with respect to the ultimate global bound, whereas the color for each bar represents the optimal subsystem.}
\label{fig:efficiency-ii}
\end{figure}

By fixing the time within the interval $t\leq 6\pi/\omega_m$, we determine which subsystem achieves higher efficiency. In Figs.~\ref{fig:efficiency-ii}(a) and (b), we depict the optimal subsystem as a function of $g_1$ and $g_2$. In Fig.~\ref{fig:efficiency-ii}(a), $g_1$ is estimated while $g_2$ is known. Interesentingly, for small values of $g_1$, no matter what $g_2$ is, the optimal subsystem is the qubit. For $g_1>0.05\omega_m$, the optimal subsystem changes to be the light. In Fig.~\ref{fig:efficiency-ii}(b), $g_2$ is estimated while $g_1$ is known. Remarkably, the light remains the optimal subsystem for all ranges of $g_1$ and $g_2$. In practice, the time over which the estimation can happen is highly limited due to imperfections such as decoherence, damping, and dephasing. By reducing the time interval to $t\leq 2\pi/\omega_m$, we repeat the above analysis to determine the optimal subsystem. In Figs.~\ref{fig:efficiency-ii}(c) and (d), we depict the optimal subsystem as a function of $g_1$ and $g_2$. In Fig.~\ref{fig:efficiency-ii}(c), we consider the case where $g_1$ is estimated and $g_2$ is known. Interestingly, compared with Fig.~\ref{fig:efficiency-ii}(a), in most of the cases, the optimal subsystem becomes the qubit. In Fig.~\ref{fig:efficiency-ii}(d), we consider estimation of $g_2$ when $g_1$ is known. Surprinsingly, compared to Fig.~\ref{fig:efficiency-ii}(b), in the regime that $g_2$ is very small, the optimal subsystem becomes the mechanical oscillator. 

This analysis shows that the optimal subsystem for estimating $g_1$ changes between the qubit and the light depending on the affordable time interval as well as the strength of the coupling, in particualr $g_1$. For estimating $g_2$, in most of the cases, the light is the optimal subsystem. Only when the affordable time interval is short, and $g_2$ is very small, the optimal subsystem becomes the mechanical oscillator.

\section{multi-parameter estimation}\label{sec:multi-parameter estimation}
In this section, we present two different multi-parameter estimations scenarios: (i) estimating only one $g_i$ by assuming the nuisance presence of the other, the so-called nuisance estimation; and, (ii) inferring both unknown parameters $g_1$ and $g_2$ simultaneously, the so-called joint estimation.

Let us first focus on the nuisance estimation of the parameters $g_1$ ($g_2$) in the presence of $g_2$ ($g_1$). To do so, we recall from the bounds in Eq. (\ref{eq:nuisanceQCRB}) that the expression that quantifies this are the diagonal elements of the inverse of the quantum Fisher information matrix
\begin{eqnarray}
(\mathcal{Q}^{-1})_{11} &:=& \frac{\mathcal{Q}_{22}}{\mathcal{Q}_{11}\mathcal{Q}_{22}-\mathcal{Q}_{12}^2},\\
(\mathcal{Q}^{-1})_{22} &:=& \frac{\mathcal{Q}_{11}}{\mathcal{Q}_{11}\mathcal{Q}_{22}-\mathcal{Q}_{12}^2}.\label{eq:nuisance-fisher}
\end{eqnarray}
\begin{figure*}
\includegraphics[scale=0.48]{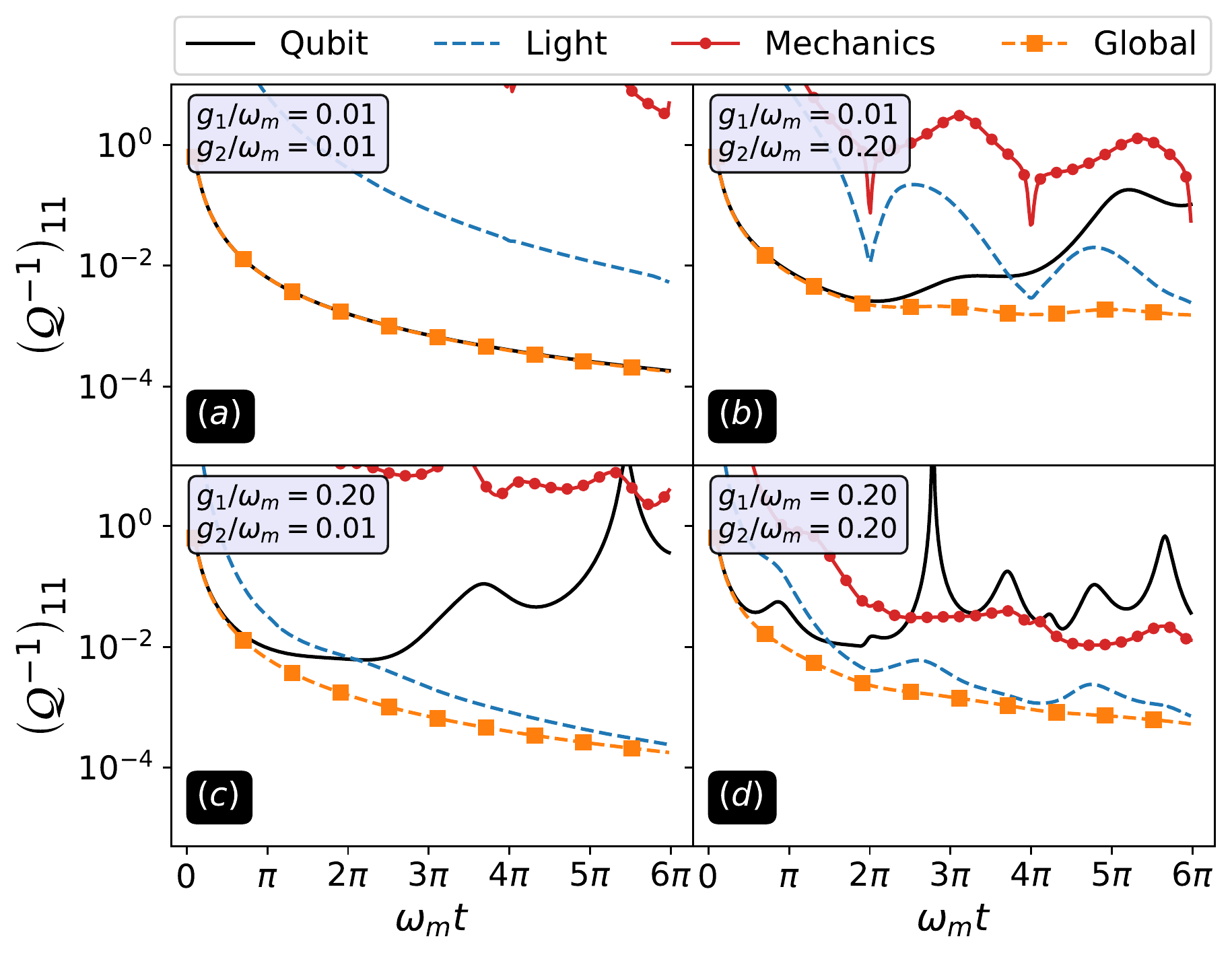}\includegraphics[scale=0.48]{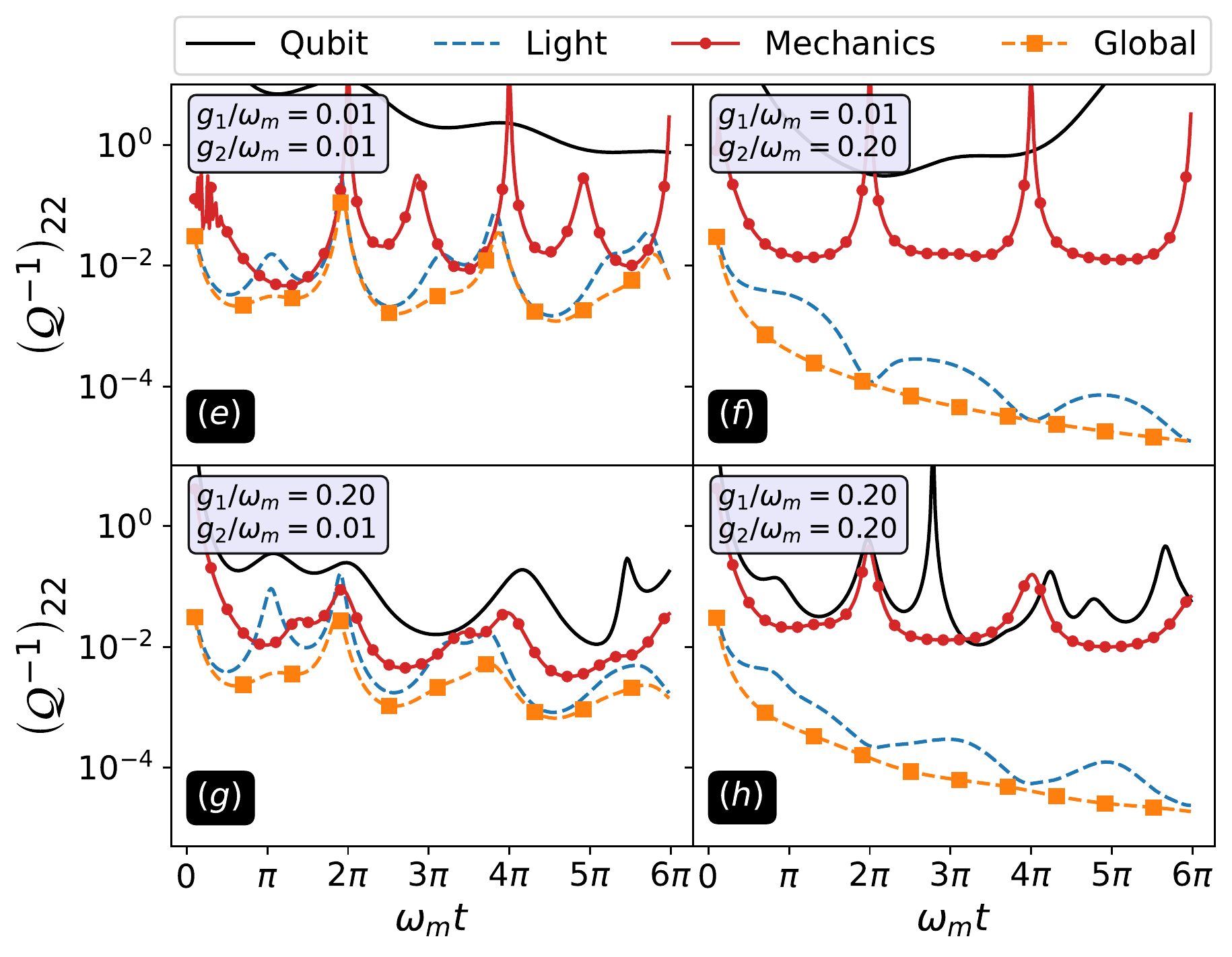}
\caption{Precision limits quantified by $(\mathcal{Q}^{-1})_{ii}$ for the estimation of $g_i$ when a nuisance parameter $g_j$ is present in the system.}
\label{fig:element-of-the-inverse}
\end{figure*}
In Fig.~\ref{fig:element-of-the-inverse}, we present the precision limits quantified by $(\mathcal{Q}^{-1})_{ii}$ for the estimation of $g_i$ when the other nuisance parameter $g_j$ is present in the system. Notably, as the figure shows, the nuisance scenario highly resembles the single-parameter estimation case already discussed and shown in Fig.~\ref{fig:single-parameter-estimation}. Hence, one concludes similar remarks, namely: (i) the mechanical oscillator (which can only encode $g_1$ using the cavity field as mediator) poorly perform for the estimation of $g_1$ [see Figs.~\ref{fig:element-of-the-inverse}(a)-(d)], (ii) analogously, the qubit system (which can only encode $g_2$ through the light field) poorly perform for the estimation of $g_2$ [see Figs.~\ref{fig:element-of-the-inverse}(e)-(h)], and (iii) having partial accessibility to the qubit or the light parties give excellent performances, even almost saturating the ultimate global bound. It is worth emphasizing that, while the above results share similar conclusions with the single-parameter scenario, they are far from being trivial. The fact that this is the case for the present model shows the relevance of determining the precision limits for each subsystem.

\begin{figure}
\includegraphics[width=0.5 \textwidth]{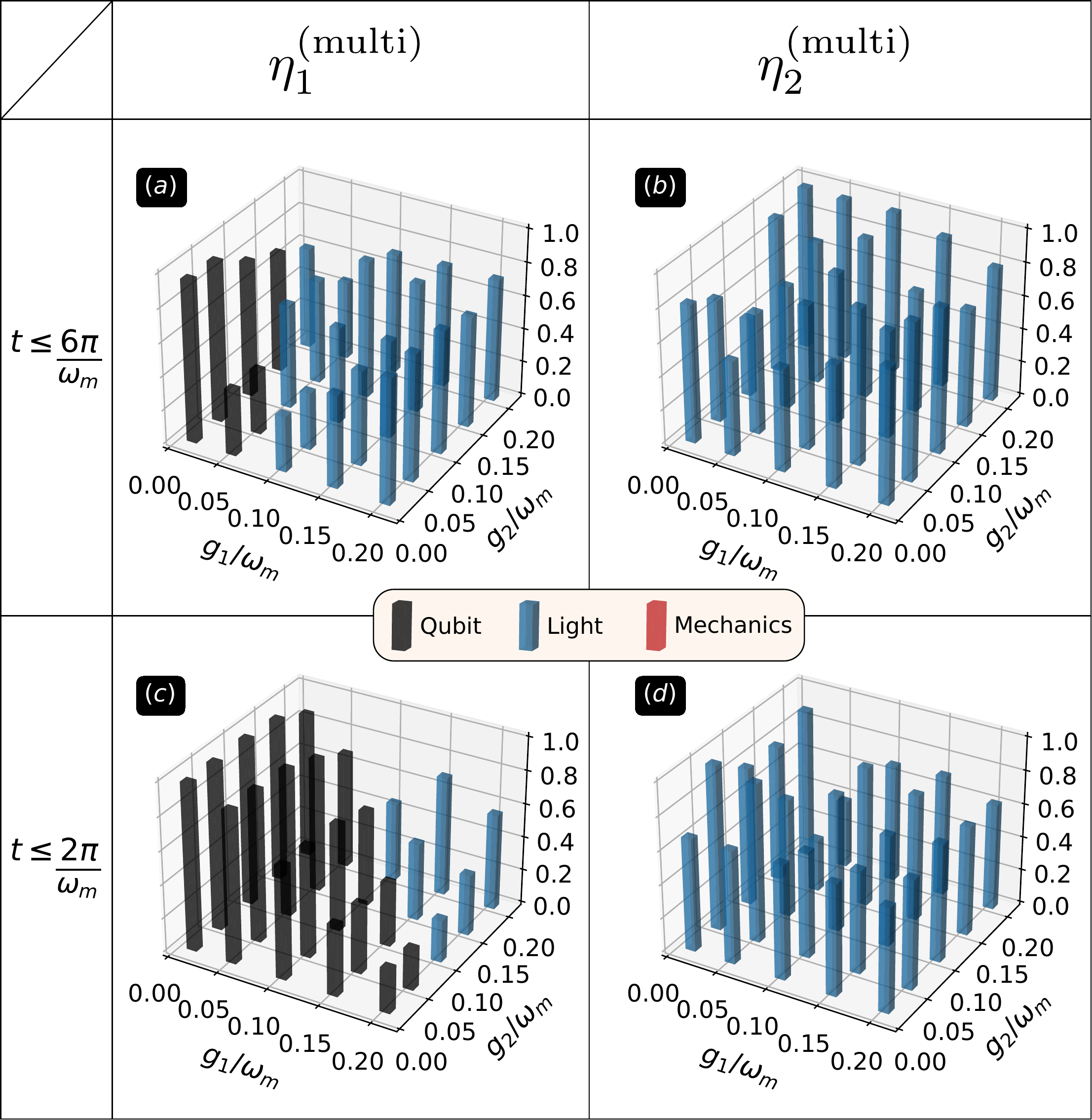}
\caption{Panels (a) and (c) show the efficiency for the estimation of $g_1$ in the presence of a nuisance unknown parameter $g_2$ for two time intervals. Similarly, panels (b) and (d) show the estimation of $g_2$ in the presence of a nuisance parameter $g_1$ for the same time windows.}
\label{fig:efficiency-nuisance}
\end{figure}

To evidence that the estimation in the presence of nuisance parameters, which employs the multi-parameter mathematical tools for its description, would still degrade the estimation of $g_i$ in the presence of an unknown $g_j$, we define its corresponding nuisance-estimation efficiency ratio as
\begin{equation}
\eta_i^{(\mathrm{multi})} = \frac{(\mathcal{Q}^{-1})_{ii, \mathrm{global}}}{(\mathcal{Q}^{-1})_{ii, \mathrm{sub}}}\Bigg\rvert_{t=t^*},\label{eq:efficiency-ratio-nuisance}
\end{equation}
where $t^*$ is the time where the optimal subsystem reaches its minimum within a given time interval. In Fig.~\ref{fig:efficiency-nuisance} we illustrate the efficiency for this scenario. As seen from the figure, even in the presence of an unknown parameter, having partial accessibility to the qubit and light subsystem still shows adequate performance. In particular, in panels (a) and (c) of Fig.~\ref{fig:efficiency-nuisance} one recovers the qubit-to-light efficiency transition shown in Fig.~\ref{fig:efficiency-ii} for a similar set of parameters $g_1$ and $g_2$. In Figs.~\ref{fig:efficiency-nuisance}(b) and ~\ref{fig:efficiency-nuisance}(d), it is evident that the light field performs better for both time windows with minor degrading when it is compared with the single-parameter scenario in Fig.~\ref{fig:efficiency-ii}. One concludes that, for the present hybrid model, the extra nuisance parameter adds a feeble noise in the final precision limits for this particular multi-parameter estimation scenario.

We now focus in setting the precision limits when both parameters are unknown, the so-called joint estimation. In particular we will consider as our figure of merit the scalar bound in Eq. (\ref{eq:scalarQCRB}), that in our scenario can be written explicitely as (in the following we wil omit the number of repetition fo the experiment $M$)
\begin{equation}
\mathrm{Var}[g_1]+\mathrm{Var}[g_2]\geq \mathrm{Tr}[\mathcal{Q}^{-1}]:=\frac{\mathcal{Q}_{11}+\mathcal{Q}_{22}}{\mathcal{Q}_{11}\mathcal{Q}_{22}-\mathcal{Q}_{12}^2},\label{eq:scalar-bound}
\end{equation}
where $\mathrm{Var}[g_i]$ is the variance for the parameter $g_i$. The above equations determines the uncertainty in estimating jointly the unknown parameters $g_1$ and $g_2$.

\begin{figure}
\includegraphics[width=\linewidth]{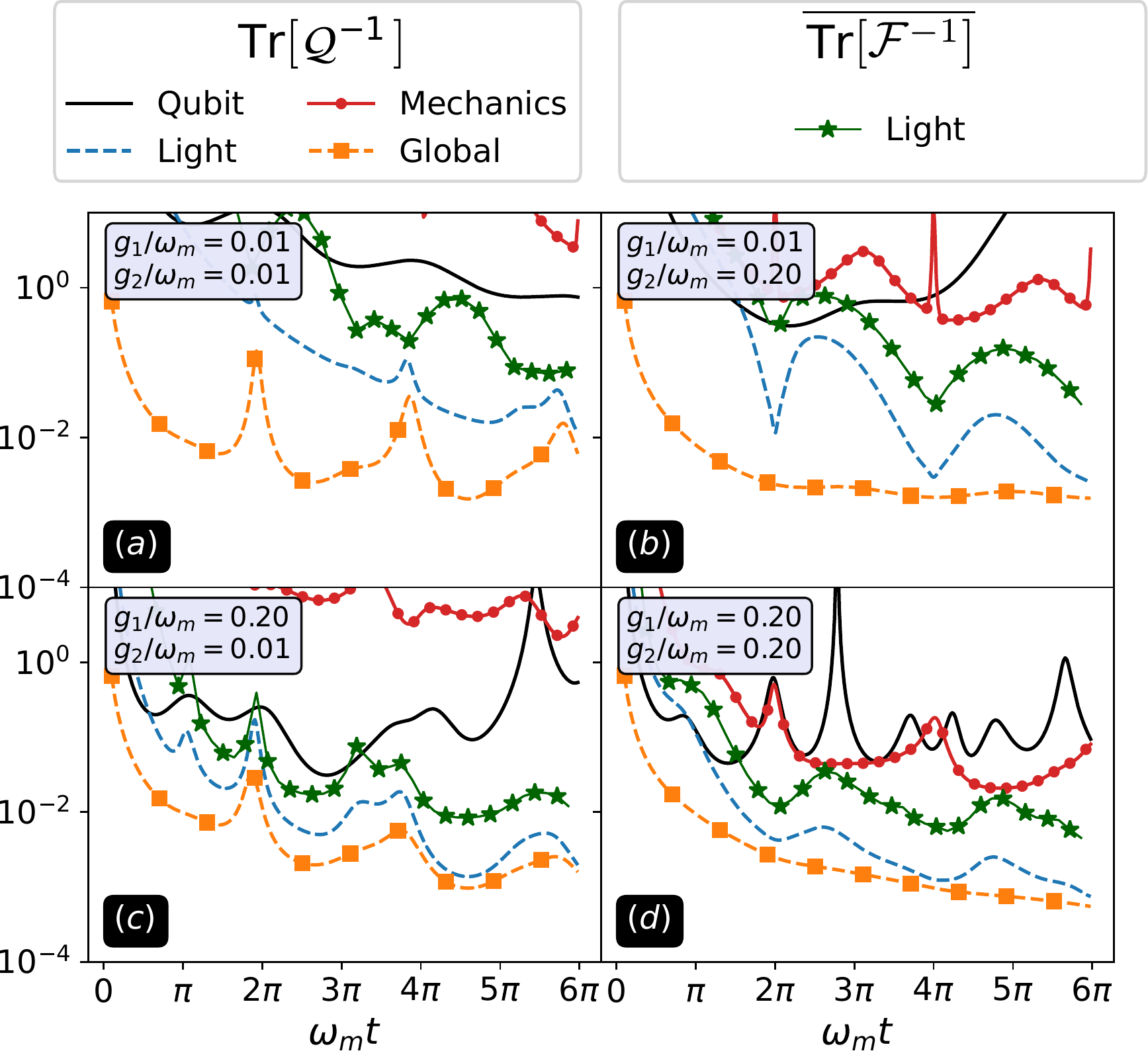}
\caption{Precision limits for the joint estimation $\mathrm{Tr}[\mathcal{Q}^{-1}]$ of parameters $g_1$ and $g_2$ as a funtion of time $\omega_mt$ for four coupling parameters. We have also included the optimized classical Fisher information scalar bound, $\overline{\mathrm{Tr}[\mathcal{F}^{-1}]}$, when the light field is measured using a homodyne detection scheme.}
\label{fig:joint-estimation}
\end{figure}

In Fig.~\ref{fig:joint-estimation} we plot the joint uncertainty, quantified by $\mathrm{Tr}[\mathcal{Q}^{-1}]$, for the simultaneous estimation of $g_1$ and $g_2$. The figure shows that both the qubit and the mechanical parties give poor performances compared to previous estimation scenarios. This can be explained by separating the joint estimation expression into its nuisances elements, i.e., $\mathrm{Tr}[\mathcal{Q}^{-1}]=(\mathcal{Q}^{-1})_{11}+(\mathcal{Q}^{-1})_{22}$. As discussed in Fig.~\ref{fig:element-of-the-inverse}, while the qubit (mechanics) provides a good performance in estimating $g_1$ ($g_2$), it fails in performing efficiently for the coupling parameter $g_2$ ($g_1$). Consequently, the overall additive operation results in estimating both coupling parameters jointly with deficient performances.  The above bad additive compensation undermines the qubit and the mechanical oscillator as good probes when the system's parameters are estimated jointly. On the other hand, as evident from Fig.~\ref{fig:joint-estimation} the light subsystem performs exceptionally well within the considered time window. Remarkably, even almost approaching the ultimate precision limits given by accessing the system globally. 

As we have mentioned in Sec. \ref{sec:parameter-estimation}, the multiparameter scalar quantum Cram\'er-Rao bound is not in general achievable. For this reason we will here analyze also the performance of a particular measurement strategy, in order to derive the corresponding classical Fisher information matrix $\mathcal{F}$ and the corresponding multiparameter scalar bound. Previous sections have studied the bounds in estimating $g_1$ and $g_2$ in a hybrid nonlinear system with partial accessibility. In particular, our results show that the information content in the cavity field makes it the most suitable subsystem to probe the coupling parameters jointly (see Fig.~\ref{fig:joint-estimation}). Therefore, we will only present a feasible measurement of the light field for the simultaneous estimation of $g_1$ and $g_2$, the widely used homodyne detection~\cite{ferraro2005gaussian}. As presented in Sec.~\ref{sec:parameter-estimation}, the scalar bound that quantifies the precision limits in estimating $g_1$ and $g_2$ simultaneously for a fixed measurement basis is:
\begin{equation}
\mathrm{Var}[g_1]+\mathrm{Var}[g_2]\geq \mathrm{Tr}[\mathcal{F}^{-1}]
\end{equation}
where the equality is reached for an optimal estimator and $\mathcal{F}$ is the classical Fisher information whose elements can be evaluated via Eq. (\ref{eq:classical_fisher_matrix}), and by considering the homodyne conditional probability $p(x_{\Phi_\mathrm{LO}}|\boldsymbol{g})$. In particular we can compute this probability as follows
\begin{equation}
p(x_{\Phi_\mathrm{LO}}|\boldsymbol{g}) = \mathrm{Tr}\left[ |x_{\Phi_\mathrm{LO}} \rangle \langle x_{\Phi_\mathrm{LO}}| \rho_\mathrm{light}(t)\right],
\end{equation}
where $|x_{\Phi_\mathrm{LO}}\rangle$ is the eigenvector of the rotated quadrature operator $x_\phi$ with local oscillator phase $\phi$ defined as:
\begin{equation}
x_{\Phi_\mathrm{LO}} = \frac{a e^{-i\Phi_\mathrm{LO}} + a^\dagger e^{i\Phi_\mathrm{LO}} }{\sqrt{2}},
\end{equation}
and $\rho_\mathrm{light}(t)$ is the reduced density matrix of the light field. Notice that, the performance of the homodyne detection depends upon the choice of the local phase $\Phi_\mathrm{LO}$. However, this phase is known and tunable in real experiments, and therefore, we optimize the homodyne detection procedure over $\Phi_\mathrm{LO}$ as \begin{equation}
\overline{\mathrm{Tr}[\mathcal{F}^{-1}]} :=\min_{-\pi\leq\Phi_\mathrm{LO}\leq+\pi} \mathrm{Tr}[\mathcal{F}^{-1}].
\end{equation}
In Fig.~\ref{fig:joint-estimation}, we contrast the classical Fisher information bound $\overline{\mathrm{Tr}[\mathcal{F}^{-1}]}$ with the quantum bound $\mathrm{Tr}[\mathcal{Q}^{-1}]$. As seen from the figure, the homodyne detection for the light field performs adequately within the time window. Interestingly, despite the fact that the simple optimized homodyne detection is not the optimal measurement basis, its performance is not very far from the optimal one. This shows that one can jointly determine $g_1$ and $g_2$ over a wide range merely by performing the homodyne detection on the cavity field.

\section{Concluding remarks}\label{sec:conclusions}

In this paper, we investigate the possibility of dynamically estimating the couplings between qubit-light and light-mechanics in a hybrid optomechanical system. Although, the quantum state of the entire system carries a wealth of information about the couplings, in practice, extracting such information demands global measurements which are not readily available. Thus, the most sensible approach is to estimate the couplings through measurements on one of the subsystems, namely the qubit, cavity light, or the mechanical oscillator. Due to the entanglement between different components of the system, the reduced density matrix of each subsystem is mixed, and thus, it is not obvious how much information one can extract via this partial accessibility. We show that, indeed the couplings can be estimated through partial accessibility with the precision not very far from the global bound. Our comprehensive analysis shows that for estimating the light-mechanics coupling, the light field is dominantly the optimal subsystem to be measured. Interestingly, this is also the case for simultaneous joint estimation of the two couplings. On the other hand, for estimating the qubit-light coupling, depending on the situation either the qubit or the light field can be the optimal subsystems. For instance, in the case of single-parameter and nuisance multi-parameter estimation, depending on the range of the qubit-light coupling, the optimal subsystem can change from the qubit to the cavity field. The reason that light is the most suitable subsystem for inferring the couplings is that the light is responsible for mediating the interaction between the other two parties. Finally, for the sake of completeness, we show that a simple widely used homodyne measurement on the light degrees of freedom, can extract the values of the couplings with a fair precision.

\section*{Acknowledgments}
A.B. acknowledges support from the National Key R\& D Program of China for Grant No. 2018YFA0306703, the National Science Foundation of China for Grants No. 12050410253 and No. 92065115, and the Ministry of Science and Technology of China for the Young Scholars National Foreign Expert Project for Grant No. QNJ2021167001L. V.M. thanks the National Natural Science Foundation of China for Grant No. 12050410251, the Chinese Postdoctoral Science Fund for Grant No. 2018M643435, and the Ministry of Science and Technology of China for the Young Scholars National Foreign Expert Project for Grant No. QNJ2021167004.

\appendix

\section{Jaynes-Cummings-like dynamics}\label{appendix-1}
This section puts forward key aspects of the tripartite system which will help us understand the role of the mechanical displacement in the Jaynes-Cummings dynamics. Indeed, several works have already considered such hybrid Hamiltonian in Eq.~\eqref{eq:H}. In particular, within the single-photon subspace, it has been studied the regimes where $g_1 \ll \omega_m, g_2$ and $g_2 \approx \omega_m$ leading to slow Rabi oscillations, and the case $g_2 \ll \omega_m$ where they are almost suppressed~\cite{PhysRevA.92.043822}. Additionally, lifting the single-photon subspace restriction has shown that the population inversion exhibits anomalous oscillations induced by the mechanical displacement for $g_2\ll \omega_m$ and different initial states for the light field and the mechanical object~\cite{Asiri_2018}. For the sake of completeness, we briefly present and discuss the derivation of the Jaynes-Cummings-like Hamiltonian derived in Refs.~\cite{PhysRevLett.112.013601, restrepo:tel-01099806}, where the mechanical displacement explicitly couples to a qubit-light polariton doublet.

Let us first consider the polariton-phonon basis $\{|\pm^{(n)}, m\rangle\}$ for the resonant case $(\omega_q{=}\omega_c)$, where $m$ is an integer and $|\pm^{(n)}\rangle$ are the polariton Jaynes-Cummings dressed states defined as: ($n{=}0$ returns $|g, 0\rangle$)
\begin{eqnarray}
\nonumber \forall n \in \mathbb N, n &\neq& 0,\\
|+^{(n)}\rangle &=& \frac{1}{\sqrt{2}}(|g,n\rangle + |e,n-1\rangle),\\
|-^{(n)}\rangle &=& \frac{1}{\sqrt{2}}(|g,n\rangle - |e,n-1\rangle).
\end{eqnarray}
As known, the polariton basis exactly diagonalizes the Jaynes-Cummings Hamiltonian (i.e., $g_2{=}0$), and hence, the basis $\{|\pm^{(n)}, m\rangle\}$ enables to write the hybrid Hamiltonian in Eq.~\eqref{eq:H} as follows:
\begin{eqnarray}
\nonumber H &=& \omega_m b^\dagger b + \sum_{n\in \mathbb{N}} H^{(n)},\\
\nonumber H^{(n)} &=& \left(n - \frac{1}{2}\right) \omega_c \mathbb{I}^{(n)} + \frac{\Omega^{(n)}}{2}\sigma_z^{(n)}\\
&-&g_2 \left\{\frac{1}{2}\sigma_x^{(n)} + \left( n - \frac{1}{2} \right) \mathbb{I}^{(n)} (b + b^\dagger ) \right\}, \label{eq:almost-polaron}
\end{eqnarray}
where $\Omega^{(n)}{=}2\sqrt{n}g_1$, $\mathbb{I}^{(n)}$ is the identity matrix in the subspace spanned by the set $\{|+^{(n)}\rangle, |-^{(n)}\rangle\}$, and $\sigma_i^{(n)}$ are Pauli matrices acting on the same polariton subspace. Two readily evident features can be drawn from Eq.~\eqref{eq:almost-polaron}, namely: (i) each of the two cavity polariton states $\{|\pm^{(n)}, m\rangle\}$ couples effectively to the mechanical's position evidenced by $\propto \sigma_x^{(n)}(b + b^\dagger)$, and (ii) each $\{|\pm^{(n)}, m\rangle\}$ contains on average $n{-}1/2$ excitations that displace the equilibrium position of the mechanical object. One can further absorbe the latter $n$-dependent mechanical equilibrium displacement by introducing
\begin{equation}
b=b_n + \frac{g_2}{\omega_m}\left(n - \frac{1}{2}\right).
\end{equation}
The above new operator $b_n$ introduces an associated Fock basis $|m^{(n)}\rangle$, with $m^{(n)} \in \mathbb{N}$ phonons for the mechanical mode centered at $\sqrt{2}g_2/\omega_m(n-1/2)$. The above transformation leads to the effective Jaynes-Cummings-like Hamiltonian for the weak single-photon optomechanical regime $g_2\ll \omega_m$~\cite{PhysRevLett.112.013601}:
\begin{equation}
H \simeq \sum_{n\in \mathbb{N}} \omega_m b_n^\dagger b_n + \sqrt{n}g_1\sigma_z^{(n)} - \frac{g_2}{2}\left(b_n^\dagger\sigma_-^{(n)} + b_n\sigma_+^{(n)} \right).\label{eq:jc-effective}
\end{equation}
It is worth emphasizing that, in the above we have performed the rotating wave approximation for each $n$ polariton subspace, namely $|\omega_m - \Omega^{(n)}|\ll \omega_m + \Omega^{(n)}$.

Since the polariton number operator $N=a^\dagger a + \sigma_+\sigma_-$ permits to diagonalize the Jaynes-Cummings Hamiltonian in the basis $\{|\pm^{(n)}, m\rangle\}$, one can employ the same techniques to diagonalize the Hamiltonian in Eq.~\eqref{eq:jc-effective} by defining a polaron number operator as
\begin{equation}
N_\mathrm{polaron}=b_n^\dagger b_n + \sigma_+^{(n)}\sigma_-^{(n)}.
\end{equation}
This readily lead us to the polaron eigenbasis
\begin{eqnarray}
\nonumber \Omega^{(n)} &\neq& \omega_m,\\
\nonumber \begin{pmatrix}
|+^{n,m^{(n)}}\rangle\\
|-^{n,m^{(n)}}\rangle\\
\end{pmatrix}&=&R(2\theta^{n,m^{(n)}})\begin{pmatrix}
|+^{(n)}\rangle|(m-1)^{n,m^{(n)}}\rangle\\
|-^{(n)}\rangle|m^{n,m^{(n)}}\rangle\\
\end{pmatrix},\\
\end{eqnarray}
where
\begin{equation}
R(2\theta^{n,m^{(n)}})=\begin{pmatrix}
\cos\theta^{n,m^{(n)}} & \sin\theta^{n,m^{(n)}}\\
\sin\theta^{n,m^{(n)}} & -\cos\theta^{n,m^{(n)}}\\
\end{pmatrix},
\end{equation}
such that
\begin{eqnarray}
\nonumber \theta^{n,m^{(n)}} &\in& \left(-\frac{\pi}{2}, 0\right],\\
\tan2\theta^{n,m^{(n)}} &=& \frac{g_2\sqrt{m^{(n)}}}{\sqrt{(\Omega^{(n) - \omega_m})^2+g_2^2m^{(n)}}},
\end{eqnarray}
and associated polaron eigenenergies:
\begin{eqnarray}
\nonumber E_\pm^{n, m^{(n)}} &=& \pm \sqrt{\frac{(\Omega^{(n)}-\omega_m)^2}{4}+m^{(n)}\frac{g_2^2}{4}}\\
&+&\omega_0^{(n)}+\left(m^{(n)} - \frac{1}{2} \right)\omega_m.
\end{eqnarray}
While the unitary dynamics can be now solved straighforward using
\begin{equation}
|\psi(t)\rangle=\sum_{ \substack{ n=0\\ m^{(n)}=0}}^\infty \sum_{j=+,-} e^{-itE_j^{n, m^{(n)}}} |j^{n,m^{(n)}}\rangle \langle j^{n,m^{(n)}}|\psi(0)\rangle,\label{eq:so-equation}
\end{equation}
special attention must be paid for the inner product between the Fock states of $b$ and the displaced basis $b_n$ which obeys
\begin{eqnarray}
\nonumber \forall m^{(n)}, l &\in & \mathbb{N},\\
\nonumber \langle l|m^{(n)}\rangle &=& L_{m^{(n)}}^{l-m^{(n)}}\left(\left[ \frac{g_2}{\omega_m}\left(n - \frac{1}{2} \right) \right]^2 \right) e^{-\left[ \frac{g_2}{\omega_m}\left(n - \frac{1}{2} \right) \right]^2}\\
&&\times \sqrt{\frac{m^{(n)}!}{l!}}\left[ \frac{g_2}{\omega_m}\left(n - \frac{1}{2} \right) \right]^{l-m^{(n)}},
\end{eqnarray}
where $L_{m^{(n)}}^{l-m^{(n)}}(x)$ is the generalized Laguerre polynomial of degree $m^{(n)}$ and index $l-m^{(n)}$.

The above brief review on the diagonalization of the hybrid tripartite Hamiltonian in polaron basis serves for two main purposes, namely (i) the effective Jaynes-Cummings-like Hamiltonian neglects a Star-like shift in the eigenenergies induced by $\sigma_x^{(n)}g_2^2(n-1/2)/\omega_m$, which it must be taken into account when the single-photon coupling enters the strong-to-moderate optomechanical regime; and (ii) the polaron picture shows that the mechanical oscillator couples independently to each polariton doublet. Similar effective Hamiltonians have been also derived by different techniques, for instance, by a displaced transformation picture~\cite{Asiri_2018, PhysRevA.91.013842} and through an operator approach~\cite{Ventura_Vel_zquez_2015}.

\section{Optimal subsystem in the presence of imperfections}\label{appendix-2}
This section presents the performance of the subsystems for the estimation of $g_1$ and $g_2$ in the presence of imperfections. For our purpose, we will restrict the analysis only to imperfections arising from the dynamics while keeping the whole measurement procedure with perfect efficiency. As known, any physical system interacts unavoidably with one or more reservoirs, an inaccessible system with larger degrees of freedom than the system of interests, that generally causes detrimental effects to the system's dynamics. From a practical perspective, investigating the system under such detrimental effects is of utmost importance as it determines its feasibility in a more real experimental scenario. To have a fair comparison between the unitary and the non-unitary dynamics, we consider our system to evolve from
\begin{equation}
\rho(0)=|g\rangle\langle g|\otimes|\alpha\rangle\langle\alpha|\otimes\sum_{n=0}^\infty \frac{\bar{n}^{n}}{(1+\bar{n})^{n+1}}|n\rangle\langle n|.
\end{equation}
As seen from the above, while both the qubit and the cavity are initialized in experimentally available pure states, we have left the mechanical party to evolve from a thermal mixed state in Fock basis and phonon number occupancy $\bar{n}$. In particular, we consider the initial state of the qubit as its ground state $|g\rangle$, the cavity field with coherent amplitude $\alpha=2$, and the mechanical oscillator with phonon mean value $\bar{n}=1$. 
\begin{figure*}[t]
\includegraphics[width=\textwidth]{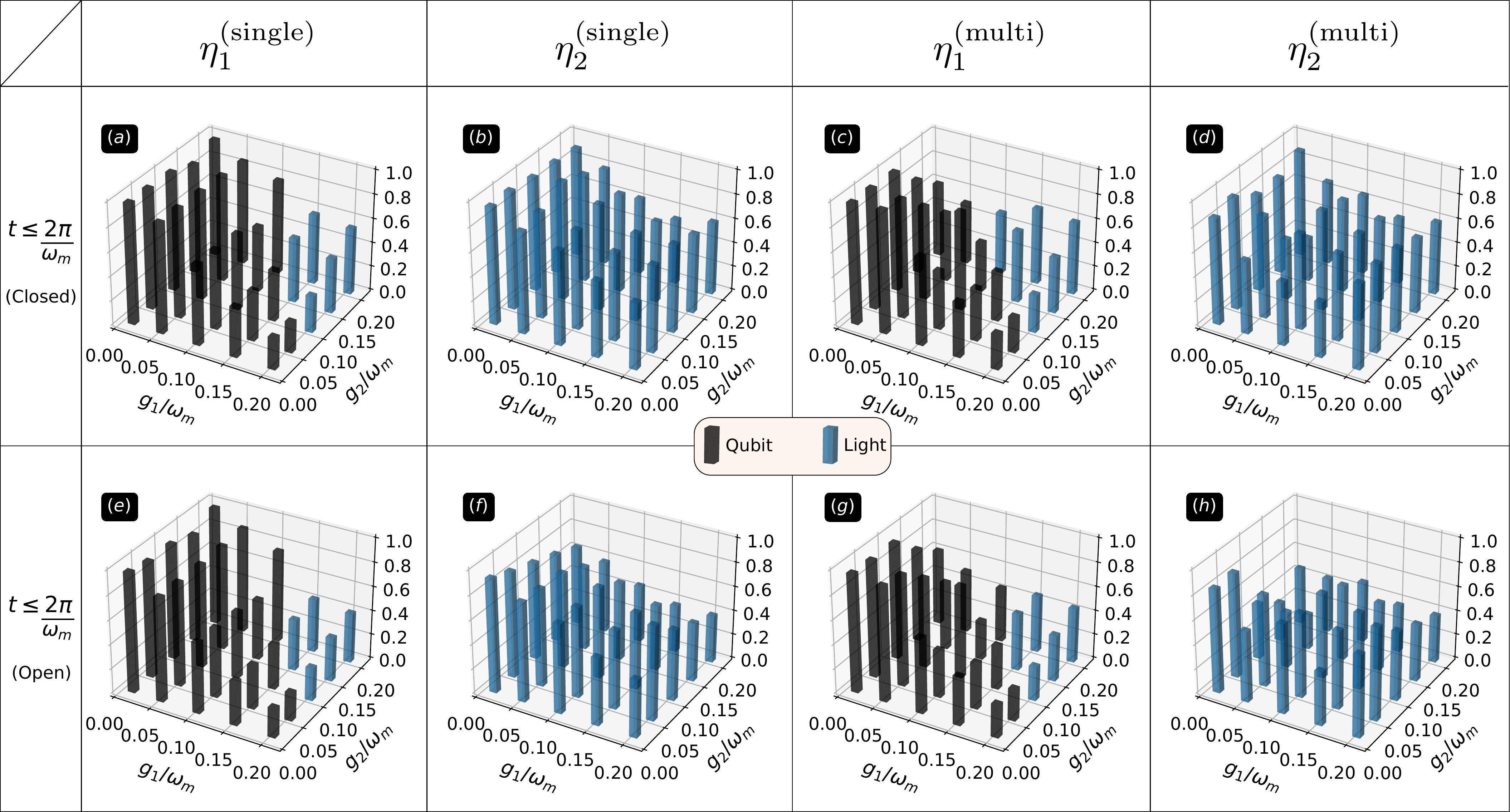}
\caption{Panels (a) to (d) show the efficiency ratios for the single-parameter and nuisance multi-parameter cases when the optimal subsystem and the global state are computed from a lossless closed dynamics. Panels (e) to (h) show the same as above when the optimal subsystem decoheres while the global state remains lossless.}
\label{fig:open-quantum-system-global-subsystem}
\end{figure*}

To model the open (non-unitary) quantum dynamics, we solve the Born-Markov master equation
\begin{multline}
\frac{d\rho}{dt}=-i[H,\rho] +\frac{\kappa}{2}\mathcal{D}[a]\rho + \\
\frac{\Gamma}{2}(1+\bar{N})\mathcal{D}[b]\rho + \frac{\Gamma}{2}\bar{N}\mathcal{D}[b^\dagger]\rho +\frac{\gamma}{4}\mathcal{D}[\sigma_z]\rho,\label{eq:master-equation}
\end{multline}
where
\begin{equation}
\mathcal{D}[O]=2O\rho O^\dagger - \rho O^\dagger O -  O^\dagger O \rho,
\end{equation}
and $\kappa, \Gamma, \gamma$ account for the cavity intensity decay rate, the mechanical damping rate, and the pure dephasing rate, respectively. In Eq.~\eqref{eq:master-equation}, $\bar{N}$ is the average phonon number in thermal equilibrium $\bar{N}=(e^{\omega_m/k_BT}-1)^{-1}$, where the Planck constant has been set to $\hbar = 1$, $T$ is the temperature of the reservoir, and $k_B$ is the Boltzmann constant. Due to the large difference between the mechanical and the cavity modes, i.e., $\omega_m \ll \omega_c$, we have omitted the average photon number in thermal equilibrium. In what follows, we consider the hybrid system to evolve under the rates $\kappa=0.01\omega_m, \Gamma=10^{-5}\omega_m, \gamma=0.01\omega_m$ embedded in a reservoir with $\bar{N}=100$ phonon excitations on average.

We first focus in quantifing how much information content is lost when the optimal subsystem decoheres, whereas the global bound remains lossless. To do so, we consider the same efficiency ratios as defined in the main body of this paper, namely, for the single-parameter estimation case
\begin{equation}
\eta_i^{(\mathrm{single})} = \frac{(\mathcal{Q}_{ii})^{-1}_\mathrm{global, closed}}{(\mathcal{Q}_{ii})^{-1}_\mathrm{sub}}\Bigg\rvert_{t=t^*},
\end{equation}
and
\begin{equation}
\eta_i^{(\mathrm{multi})} = \frac{(\mathcal{Q}^{-1})_{ii, \mathrm{global, closed}}}{(\mathcal{Q}^{-1})_{ii, \mathrm{sub}}}\Bigg\rvert_{t=t^*},
\end{equation}
for the nuisance multi-parameter estimation scenario. Notice that, we have stressed in the above numerators that the global state evolves as a closed lossless system, while the optimal subsystem will undergo a closed or open dynamics.
\begin{figure*}[t]
\includegraphics[width=\textwidth]{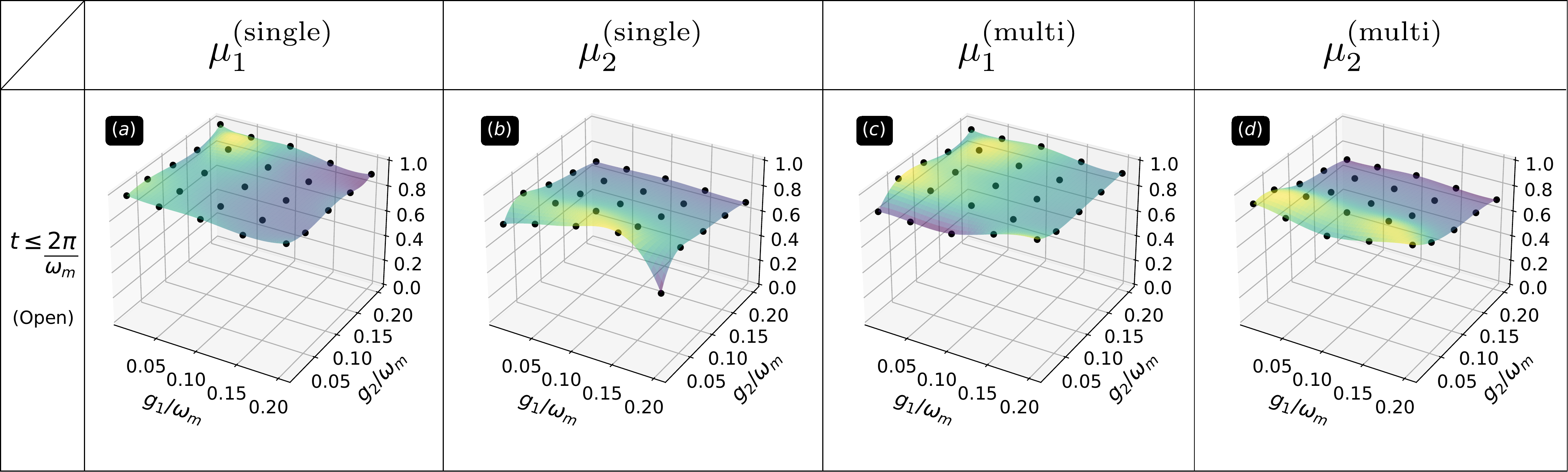}
\caption{Panels (a) and (b) show the single-parameter efficiency ratio between closed and open global states. Panels (c) and (d) show the nuisance multi-parameter efficiency ratio between global states.}
\label{fig:open-quantum-system}
\end{figure*}

In Figs.~\ref{fig:open-quantum-system-global-subsystem}(a)-(d), we show the efficiency ratios for the single-parameter and nuisance multi-parameter cases when the optimal subsystem and the global state are computed from a lossless closed dynamics. Notice that, with the choice of a more experimental mechanical state initialization, the ground state with $\bar{n}\approx 1$~\cite{OConnell2010, Chan2011}, one reaches similar conclusions as to when the system evolves from a coherent mechanical oscillator. In particular, from Figs.~\ref{fig:open-quantum-system-global-subsystem}(a)-(d), it is evident that for estimating $g_1$ the optimal subsystems can be either the qubit or the cavity field depending on the set of $g_1$ and $g_2$ parameters. Moreover, very high performances can be reached, especially for weak values of $g_1$. Additionally, for the estimation of $g_2$, the cavity field is the dominant optimal subsystem for all the considered range of $g_1$ and $g_2$ parameters. In Figs.~\ref{fig:open-quantum-system-global-subsystem}(e)-(h), we show the efficiency ratios for the single-parameter and nuisance multi-parameter cases when the optimal subsystem decoheres and the global state remains to evolve losslessly. Interestingly, as seen from the figures, the efficiency ratios are mildly attenuated. In other words, for the set of lossy parameters considered here, one can argue that not much information content is lost within that time interval, making the single-parameter and the nuisance multi-parameter estimation robust under decoherence.

We now turn the attention to quantify how much information content the global state loses in the presence of decoherence. To do so, let us define the single-parameter efficiency ratio between global states as
\begin{equation}
\mu_i^{(\mathrm{single})} = \frac{(\mathcal{Q}_{ii})^{-1}_\mathrm{global, closed}}{(\mathcal{Q}_{ii})^{-1}_\mathrm{global, open}}\Bigg\rvert_{t=t^*},
\end{equation}
and
\begin{equation}
\mu_i^{(\mathrm{multi})} = \frac{(\mathcal{Q}^{-1})_{ii, \mathrm{global, closed}}}{(\mathcal{Q}^{-1})_{ii, \mathrm{global, open}}}\Bigg\rvert_{t=t^*},
\end{equation}
being the nuisance multi-parameter efficiency ratio between global states undergoing a closed and an open evolution.

In Figs.~\ref{fig:open-quantum-system}(a)-(d), we plot the single-parameter efficiency ratio between global states as well as the nuisance multi-parameter efficiency ratio between global states. As seen from the figures, the attenuation of the information content in the whole state is around $\sim 20\%$ for the set of damping ratios considered for the numerical simulations. One can conclude that both the optimal subsystems as well as the global bounds in the presence of imperfections still have enough information for the estimation of the $g_1$ and $g_2$ parameters.

\bibliographystyle{apsrev4-1}
\bibliography{Polaron_v4}

\end{document}